\documentclass{jfm}
\usepackage{graphicx}
\usepackage{epstopdf, epsfig}
\usepackage{tikz}

\usepackage{verbatim}
\usepackage{graphicx}
\usepackage{amsmath}%
\usepackage{longtable}%
\usepackage{bm}%
\usepackage{textcomp}
\usepackage{color}
\usepackage{hyperref}

\usepackage[normalem]{ulem}

\definecolor{mycol1}{rgb}{0.9047,0.1918,0.1988}
\definecolor{mycol2}{rgb}{0.2941,0.5447,0.7494}
\definecolor{mycol3}{rgb}{0.3718,0.7176,0.3612}
\definecolor{mycol4}{rgb}{1.0000,0.5482,0.1000}
\definecolor{mycol5}{rgb}{0.8650,0.8110,0.4330}
\definecolor{mycol6}{rgb}{0.6859,0.4035,0.2412}
\definecolor{mycol7}{rgb}{0.9718,0.5553,0.7741}
\definecolor{mycol8}{rgb}{0.6400,0.6400,0.6400}
\definecolor{mycol9}{rgb}{0.6365,0.3753,0.6753}

\shorttitle{Mach-number-dependent dissipative anomaly}
\shortauthor{S. Alam et al.}
\title{Mach-number-dependent dissipative anomaly in isothermal compressible turbulence} 

\author{Shadab Alam\aff{1}, Georgy Zinchenko\aff{1}, Christoph Federrath\aff{2} 
and \\J\"org Schumacher\aff{1}
\corresp{\email{ joerg.schumacher@tu-ilmenau.de}}
}
\affiliation{
\aff{1} Institut f\"ur Thermo- und Fluiddynamik, Technische Universit\"at Ilmenau, Postfach 100565, D-98684 Ilmenau, Germany
\aff{2} Research School of Astronomy and Astrophysics, Australian National University, Canberra, ACT 2611, Australia
}

\begin{document}
\maketitle

\begin{abstract}
Using a comprehensive set of three-dimensional, high-resolution direct numerical simulations, we investigate the existence of a dissipative anomaly in isothermal, homogeneous, isotropic compressible turbulence driven by solenoidal forcing. We find that the total kinetic-energy dissipation rate, as well as its solenoidal and dilatational components, approaches finite asymptotic values with increasing Reynolds number $Re$. The normalized mean dissipation rates collapse onto two distinct branches: one corresponding to the subsonic and transonic regimes, with root-mean-square Mach numbers ($M_{\rm rms}\lesssim 1$), and another to the highly supersonic regime, with ($M_{\rm rms}\ge 3$). This two-branch Mach-number dependence is most pronounced for the total kinetic-energy dissipation. For the solenoidal and dilatational dissipation rate components, the dependence on $Re$ depends in addition on the specific choice of the integral scale and root-mean-square velocity. Despite grid resolutions of up to $2048^3$ points the Reynolds numbers accessible are not sufficiently large to distinguish conclusively between a weak and a strong dissipative anomaly. We substantiate these findings using three complementary approaches: (i) a detailed analysis of the mechanisms responsible for dissipation generation, based on the corresponding dissipation-rate balance equations and their individual production terms; (ii) an investigation of precursors of anomalous dissipation using the Duchon--Robert framework extended to compressible flows; and (iii) a geometrical characterization of regions of intense dissipation. Taken together, these analyses provide consistent evidence for the existence of a dissipative anomaly in isothermal compressible turbulence, while leaving its precise weak or strong character unresolved.
\end{abstract}


\section{Introduction}
\label{sec:intro}
The dissipation of kinetic energy lies at the heart of the turbulent energy cascade. Energy injected at large scales is transferred through a hierarchy of progressively smaller scales, where it is ultimately dissipated by molecular viscosity. Despite the direct role of viscosity in this process, turbulent flows exhibit the remarkable property that, away from boundaries, the mean kinetic-energy dissipation rate becomes independent of the fluid viscosity when the latter is sufficiently small, or equivalently, when the Reynolds number is sufficiently large. This asymptotic viscosity independence is commonly referred to as the dissipative anomaly. Compelling early experimental evidence for this behavior was provided by \citet{sreeni1984} for {\em incompressible} turbulence. Their grid-turbulence measurements demonstrated that the normalized mean dissipation rate approaches a constant at sufficiently large Reynolds numbers, i.e.
\begin{equation}
    \frac{\langle\epsilon\rangle }{u_{\rm rms}^{3}/L} ={\rm constant}.
    \label{DA0}
\end{equation}
Here, $\epsilon({\bm x},t)$ is the kinetic-energy dissipation rate, $u_{\rm rms}=\sqrt{\langle{\bm u}^{2}\rangle}$ is the root-mean-square (rms) velocity of the turbulent velocity field ${\bm u}({\bm x},t)$, with $\langle \cdot \rangle$ denoting spatial and temporal averaging, and $L$ is the integral length scale,
\begin{equation}
L=\frac{3\pi}{4}\frac{\displaystyle\int_{0}^{\infty} k^{-1}E(k){\rm d}k}{\displaystyle\int_{0}^{\infty} E(k){\rm d}k},
\label{eq:int_length}
\end{equation}
where $E(k)$ is the energy spectrum and $k$ is the wavenumber magnitude \citep{SSY07}. Several subsequent experiments \citep{Pearson2002,Puga2017} and direct numerical simulations (DNS) of periodic box turbulence  \citep{sreeni98b,Kaneda2003,Bos2007,McComb2015,Kitamura2025} have provided further support for, and insight into, this asymptotic behavior.

A theoretical basis for this seemingly paradoxical phenomenon was proposed much earlier by \citet{onsag1949}. He related the persistence of dissipation in the inviscid limit to the roughness of the velocity field and argued that anomalous dissipation may occur for sufficiently irregular weak solutions of the Euler equations, corresponding to H\"older exponents $h<1/3$; see also \cite{Dubrulle2019,Eyink_2024,Sreenivasan:ARCMP2024} and \cite{Eyink2026} for reviews. The convex-integration approach of \citet{LS2009,LS2014} led to the construction of anomalously dissipative weak Euler solutions for $h<1/10$, and \citet{Isett2018} later reached the Onsager threshold, establishing the existence of such solutions for $h<1/3$.

This link between the roughness of the velocity field and anomalous dissipation can also be understood in terms of interscale kinetic energy transfer. \citet{DR2000} formulated this connection through a local energy balance based on spatial coarse-graining and introduced an energy-defect term that depends entirely on velocity increments and can remain finite as the coarse-graining scale tends to zero. They further related this defect to third-order velocity increments through Kolmogorov's $4/3$-law. \citet{Eyink_2003} related the same energy-defect term to the $4/5$- and $4/15$-laws. Recently, \citet{Zinchenko_2024} applied the Duchon--Robert formulation to Burgers-vortex type models and DNS of homogeneous isotropic turbulence, the latter of which obey Burgers-like vortex stretching. They associated enhanced local dissipation with intense vortex-stretching events and identified these structures as local precursors to anomalous dissipation at finite Reynolds numbers. 

In compressible turbulence, the energy cascade is modified by thermodynamic fluctuations, which alter the interscale energy flux, and by dilatational motions, which introduce additional source- or sink-like contributions to the scale-by-scale energy balance \citep{Galtier_2011}. The kinetic-energy cascade, however, remains dominated by local interactions, and can proceed conservatively at smaller scales if the conversion between kinetic and internal energy through pressure--dilatation is limited to larger scales \citep{aluie2011comp}. Numerical evidence for such a conservative cascade, with substantial contributions from compressive motions, was reported in \citet{AluieAJF2012,WYS+2013}, and the role of compressive motions increases with Mach number \citep{wang2018kinetic,HDMR2026}. Building on Onsager's approach, \citet{ED2018} extended the theory of anomalous dissipation to compressible turbulence and identified two mechanisms for anomalous kinetic-energy dissipation: the local energy cascade and a pressure-work defect. Irreversible heating across stationary planar shocks provided an explicit example of the latter mechanism. More recently, \citet{zinchenko2026} extended the Duchon--Robert formulation to compressible turbulence and identified additional contributions to anomalous dissipation from compressibility. The application of this framework to one-dimensional shock-forming gas dynamics flows revealed intense anomalous dissipation localized at shock fronts.

Numerical investigations of the dissipative anomaly in compressible turbulence comprise only a few studies, which have so far been carried out in the subsonic regime \citep{JD2016,JDSJFM2021}. They reported that the classical incompressible scaling persists under solenoidal forcing, with increasing departures as the dilatational component of the forcing increases. It has further been suggested that a dissipative anomaly in the dilatational dissipation field is likely when the solenoidal and dilatational time scales become comparable \citep{JDSJFM2021}. At high Mach numbers, even under solenoidal forcing, compressibility profoundly enriches the small-scale dynamics by  promoting increasingly strong dilatational motions and highly intermittent, shock-dominated dissipation \citep{AFS2026}. The growing influence of these compressible effects raises the question of whether the classical incompressible scaling observed for the total and solenoidal dissipation in solenoidally forced subsonic turbulence carries over to the highly supersonic regime, and whether the dilatational dissipation exhibits a dissipative anomaly? 

Addressing these questions forms the central focus of the present work. We therefore examine the total kinetic energy dissipation rate field and its solenoidal and dilatational components for a wide range of Mach numbers, ranging from nearly incompressible to highly supersonic regimes using our DNS record of three-dimensional solenoidally forced, isothermal, homogeneous isotropic turbulence \citep{AFS2026}. We show that the rescaled total mean kinetic energy dissipation rate, cf. \eqref{DA0},  tends to saturate to constant values along two branches, one for Mach numbers in the subsonic and transonic regimes and one for the supersonic regime. This is a clear indication for a Mach-number-dependent dissipative anomaly. In order to better understand this empirical result, we combine detailed investigations from three different directions: (1) We apply an extension of the Duchon--Robert framework to compressible flow by \citet{zinchenko2026} to examine the local structure of precursors to anomalous dissipation, which are connected with shock layers for the highest Mach numbers. (2) This is combined with a comprehensive geometric analysis of the high-dissipation events for total energy dissipation and in terms of its solenoidal and dilatational components. For the highest Mach numbers, the maxima of all dissipation fields become two-dimensional (2D) and therefore sheet-like, which is confirmed by box-counting and multifractal analyses. (3) We examine in detail the local generation mechanisms of high-amplitude dissipation events. A better understanding of generation and geometry of the energy dissipation fields in this setting of compressible turbulence by the analysis of DNS can also provide useful input for mathematically rigorous analysis of the dissipative anomaly in compressible turbulence, e.g.~by convex integration techniques \citep{LS2014}.    

The remainder of the paper is organized as follows. The numerical methodology and the compressible Duchon--Robert framework are introduced in \S~\ref{sec:methods}. In \S~\ref{sec:dissi_anomaly}, we investigate the dissipative anomaly and the underlying nonlinear production mechanisms, followed by an analysis of local anomalous dissipation and its relation to shock-layer structures within the compressible Duchon--Robert framework. We then extend the analysis to characterize the spatial organization and intermittency of the dissipation fields. In §~\ref{sec:multi}, these properties are examined through a geometrical analysis of the total, solenoidal and dilatational dissipation fields. Finally, the main findings are summarized in \S~\ref{sec:con} together with an outlook to future extensions.

\section{Methods}
\label{sec:methods}

\subsection{Direct numerical simulations and model parameters}
The dynamics of an {\em isothermal} compressible fluid is governed by the conservation of mass and momentum; a pressure that is directly proportional to the mass density closes the system of equations. They are given by 
\begin{align}
\frac{\partial \rho}{\partial t}
+{\bm \nabla}\cdot(\rho{\bm u}) &= 0,
\label{eq:mass}\\
\frac{\partial(\rho{\bm u})}{\partial t}
+{\bm \nabla}\cdot(\rho{\bm u}\otimes{\bm u})
&=-{\bm \nabla}p+{\bm \nabla}\cdot{\bm \sigma}
+\rho{\bm f},
\label{eq:mom}\\
p &= c_s^2 \rho\,.
\label{eq:eos}
\end{align} 
Here, ${\bm u}$ is the velocity field, $\rho$ is the mass density, $p$ is the pressure, $c_s$ is the constant speed of sound and ${\bm f}$ is the external forcing. We consider a Newtonian fluid under the Stokes hypothesis, for which the viscous stress tensor is
\begin{equation}
{\bm \sigma} =2\mu\left[{\bm S} -\frac{1}{3}({\bm \nabla}\cdot{\bm u}) {\bm I}\right] =2\mu{\bm D},
\end{equation}
where
\begin{equation}
{\bm S} =\frac{1}{2}\left[{\bm \nabla}{\bm u} +({\bm \nabla}{\bm u})^{T}\right] \qquad {\rm and} \qquad {\bm D} ={\bm S}-\frac{1}{3}({\bm \nabla}\cdot{\bm u}){\bm I}, 
\end{equation}
are the symmetric strain-rate and symmetric deviatoric strain-rate tensors, respectively, while ${\bm I}$ is the identity tensor and $\mu=\rho\nu$ the dynamic viscosity.

We solve the three-dimensional governing equations by DNS in a triply periodic cubic box using a modified version of the \texttt{FLASH} code, based on release~4.0.1 \citep{Fryxell:2000,Dubey:2008}, which employs the MUSCL--Hancock HLL5R scheme \citep{Bouchut:2010,Waagan:2011} and hybrid-precision computation \citep{Federrath:2021,FederrathOffner2025}. Turbulence is sustained by a purely solenoidal stochastic forcing ${\bm f}$ based on an Ornstein-Uhlenbeck process \citep{Federrath:2010,FederrathEtAl2022ascl}. The forcing acts at large scales over the wavenumber range $1<k L_{\rm box}/2\pi<3$, with a parabolic spectrum peaking at $k L_{\rm box}/2\pi=2$, where $L_{\rm box}$ is the side length of the cubic box. This gives a characteristic forcing length scale $L_f\simeq L_{\rm box}/2$. The isothermal compressible turbulence is homogeneous and locally isotropic. 

Our DNS cover two orders of magnitude in root mean square (rms) Mach number, $0.1\leq M_{\rm rms}\leq10$, spanning nearly incompressible to highly compressible turbulence, and a Reynolds-number range $100\lesssim Re\lesssim2400$ at each $M_{\rm rms}$ to characterize the Reynolds-number dependence of the dissipation rate. The Mach and Reynolds numbers are based on the rms velocity and forcing length scale and are given by 
\begin{equation}
M_{\rm rms}=\frac{u_{\rm rms}}{c_s}, \qquad Re=\frac{u_{\rm rms}L_f}{\nu}. 
\end{equation}
The kinematic viscosity $\nu=\mu/\rho$ is constant in each simulation, but takes different values in the simulations. All simulations resolve scales well below the Kolmogorov length, with $k_{\rm max}\eta>10$, where $\eta=(\langle\rho\rangle\nu^3 /\langle\epsilon\rangle)^{1/4}$ is the Kolmogorov length scale and $k_{\rm max}=\pi N/L_{\rm box}$ is the Nyquist wavenumber, with $N$ being the number of grid points along each coordinate direction. We list the complete simulation parameters in Appendix~\ref{appen:numerical}. A comprehensive description and validation of the numerical method for a number of statistical quantities is found in \citet{AFS2026}.

\subsection{Framework by Duchon and Robert}
Here, we briefly summarize the extension of the Duchon-Robert (DR) framework to compressible Navier-Stokes flow, which was developed in \citet{zinchenko2026} and will be used here to analyse the high-amplitude dissipation regions as precursors to anomalous dissipation. Introducing the density-weighted velocity
\begin{equation}
w_i(\bm{x},t)=\sqrt{\rho(\bm{x},t)}\,u_i(\bm{x},t),
\end{equation}
the local kinetic energy is $E=w_i w_i/2$. In terms of $w_i$, the corresponding compressible Navier-Stokes equation can be written in vector notation as
\begin{equation}
\partial_t \bm{w}
+\bm{\nabla}\cdot\left(\bm{w}\otimes\bm{u}\right)
-\frac{1}{2}\bm{w}\theta=
\frac{1}{\sqrt{\rho}}\bm{\nabla}\cdot\bm{\sigma}
-\frac{1}{\sqrt{\rho}}\bm{\nabla}p,
\qquad \theta=\bm{\nabla}\cdot\bm{u}.
\end{equation}
Following DR, the fields are regularized at coarse-graining scale $r$ by convolution with a smooth kernel $\varphi^r$,
\begin{equation}
\bm w^r=\varphi^r * \bm w=\int_V\varphi^r(\bm \xi )\bm w(\bm x +\bm \xi)d\bm \xi, \quad \varphi^r({\bm \xi})=\frac{1}{r^3}\varphi\left(\frac{{\bm \xi}}{r}\right).
\end{equation}
Combining the regularized and original momentum equations, summing over all spatial directions, and taking the limit $r\to0$, one obtains the balance of the turbulent kinetic energy density $E({\bm x},t)={\bm w}^2({\bm x},t)/2$, which is given by
\begin{equation}
\partial_t E+\bm{\nabla}\cdot(\bm{u}E)=
\bm{u}\cdot(\bm{\nabla}\cdot\bm{\sigma})
-\bm{u}\cdot\bm{\nabla}p
-\lim_{r\to0}\left(D_{wwu}+D_{w\chi p}-D_{w\chi\sigma}+\mathcal{D}^r\right).
\label{eq:DR_balance}
\end{equation}
We denote $\chi=1/\sqrt{\rho}$ and $\bm \lambda=\bm \nabla\chi$. Spatial increments are denoted for example as $\delta_{\xi} \bm w=\bm w({\bm x}+{\bm \xi})-\bm w({\bm x})$ for the field $\bm w$. The simulation domain is denoted as $V$. The additional dissipation terms $D$ on the right-hand side of \eqref{eq:DR_balance} are defined as follows, 
\begin{subequations} 
\begin{align} 
D_{wwu}(r, {\bm x}) &= \frac{1}{4} \int_V \frac{\partial \varphi^r (\bm{\xi})}{\partial \xi_j} \delta_{\xi} u_j \lvert\delta_{\xi}\bm{w}\rvert^2\; d^3 \bm{\xi}\label{eq:diss_deform}\\
D_{w\chi p}(r, {\bm x}) &= \frac{1}{2} \int_V \frac{\partial \varphi^r (\bm{\xi})}{\partial \xi_j} \delta_\xi w_j \delta_\xi \chi \delta_\xi p\; d^3 \bm{\xi},\label{eq:diss_barop}\\ 
D_{w\chi \sigma}(r, {\bm x}) &= \frac{1}{2} \int_V \frac{\partial \varphi^r (\bm{\xi})}{\partial \xi_j} \delta_\xi w_i \delta_\xi \chi \delta_\xi \sigma_{ij}\; d^3 \bm{\xi}\label{eq:diss_visc}.  
\end{align} 
\label{eq:diss_terms}
\end{subequations}
We apply the Einstein summation convention. The term $\mathcal{D}^r$ collects defects, that couple fields with their gradients. These terms do not necessarily vanish as $r\to0$. In particular, three contributions appear to exist for $\mathcal{D}^r$,
\begin{equation} 
\label{eq:diss_deffect} 
\begin{aligned} \mathcal{D}^r&=\mathcal{D}^r_{\sigma}+\mathcal{D}^r_{\theta}+\mathcal{D}^r_{\lambda}\\ &=\frac{\partial}{\partial \xi_j}[\left( \chi\sigma_{ij} w_i \right)^r - \chi \left( \sigma_{ij} w_i \right)^r] + [\left( w_i\sigma_{ij} \right)^r \lambda_j - w_i \left( \sigma_{ij} \lambda_j \right)^r] \\&+ \frac{1}{2} [w_i \left( w_i \theta \right)^r-\left( w_i w_i \right)^r \theta] + [w_i \left( p \lambda_i \right)^r- \left(w_i p \right)^r \lambda_i].
\end{aligned} 
\end{equation}
These terms \eqref{eq:diss_terms}, \eqref{eq:diss_deffect} represent further terms of energy dissipation in the compressible DR framework. The first two terms, $D_{wwu}$ and $D_{w\chi p}$, will be related to anomalous dissipation. The third term \eqref{eq:diss_visc} is an additional compressible contribution to viscous dissipation. Terms $\mathcal{D}^r$ determine conversion or transport defects that appear due to unresolved covariance with $\theta$, $\lambda$, and $\sigma_{ij}$. For finite viscosity, fields are smooth below the Kolmogorov length $\eta$. For $r<\eta$, one has $\delta_\xi f\sim|\xi|$ so that each of the three increment integrals scales like
\[
\int \frac{\partial \varphi^r}{\partial \xi} \,(\delta_\xi f)^3 \, d^3\xi \sim r^2\to 0.
\]
The same scale law is valid for the defects $\mathcal{D}^r$. In the dimensionless form, the local kinetic energy balance becomes
\begin{equation}
\begin{aligned}
    \partial_t E&+\bm{\nabla}\cdot(\bm{u}E)=
\frac{1}{Re}\bm{u}\cdot(\bm{\nabla}\cdot\bm{\sigma})
-\frac{1}{\gamma M^2_{\rm rms}}\bm{u}\cdot\bm{\nabla}p\\
&-\lim_{r\to0}\left[D_{wwu}+\frac{1}{\gamma M^2_{\rm rms}}D_{w\chi p}-\frac{1}{Re}D_{w\chi\sigma}+\mathcal{D}^r\right].
\label{eq:DR_balance_dimensionless}
\end{aligned}
\end{equation}
The polytropic index $\gamma=1$ for the isothermal case. In the incompressible limit, this balance reduces to the original form in \citet{DR2000},
\begin{equation}
\begin{aligned}
\partial_t E&+\bm{\nabla}\cdot(\bm{u}E)=-\bm{u}\cdot\bm{\nabla}p + 
\frac{1}{Re}\bm{u}\cdot(\bm{\nabla}\cdot\bm{\sigma})
-\lim_{r\to0} D_{wwu}.
\label{eq:DR_balance_dimensionless}
\end{aligned}
\end{equation}

We mention that there exists an alternative framework by \citet{Aluie2013}. It is developed for the analysis of kinetic energy transfer across scales in compressible turbulence by using coarse-graining or Favre filtering \citep{Favre1969}, e.g., $\tilde u_i=(\rho u)_i^r/\rho^r$ for a velocity component $u_i$. In this formulation, the subgrid-scale kinetic-energy flux consists of two important contributions, the deformation work $\Pi_r$ and the baropycnal work $\Lambda_r$, which are given by
\begin{equation}
\Pi^r=-\rho^r\widetilde{\tau}^r(u_i,u_j)\partial_j\tilde{u}_i \quad\mbox{and}\quad
\Lambda^r=\frac{1}{\rho^r} \overline{\tau}^r(\rho,u_j)\partial_j p^r\,,
\end{equation}
where $\widetilde{\tau}^r(u_i,u_j)=\widetilde{u_i u_j}-\tilde u_i \tilde u_j$ is the subgrid stress and $\overline{\tau}^r(\rho,u_j)=(\rho u_j)^r- \rho^r  u_j^r$ is the subgrid mass flux. The deformation term represents the work of the resolved strain against the subgrid stress, and the baropycnal term represents the work of the resolved pressure-gradient force on the subgrid mass flux. In the Duchon-Robert formulation, the terms $D_{wwu}$ and $D_{w\chi p}$ describe the corresponding deformation and baropycnal energy-transfer channels, respectively. A detailed comparison of both frameworks was provided in \citep{zinchenko2026}. Later in section \ref{sec:production}, we will thus refer to $D_{wwu}$ and $D_{w\chi p}$ as the deformation and baropycnal terms, respectively. 

\section{Empirical evidence for dissipative anomaly}
\label{sec:dissi_anomaly}
\begin{figure}
	\begin{center}
	\includegraphics[width=0.95\linewidth]{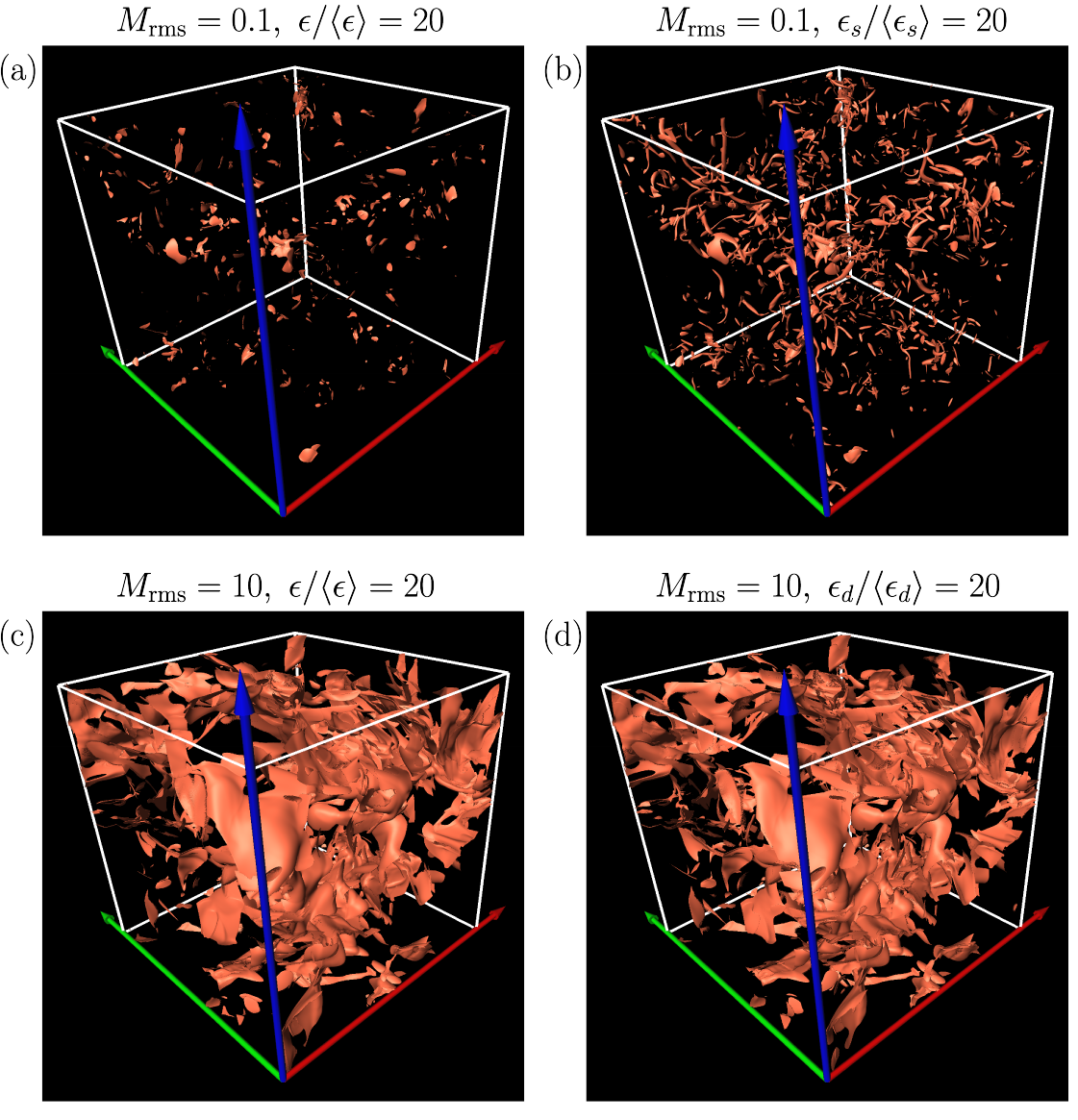}
	\end{center}
	\caption{Instantaneous isosurfaces of the decadic logarithms of the total, solenoidal and dilatational dissipation fields, $\epsilon({\bm x},t_0)$, $\epsilon_s({\bm x},t_0)$, and $\epsilon_d({\bm x},t_0)$, for the case with the highest Reynolds number $Re=2400$ and $M_{\mathrm{rms}}=0.1,\,10$. Panels $({\rm a})$ and $({\rm b})$ show isosurfaces at $\epsilon/\langle\epsilon\rangle=20$ and $\epsilon_s/\langle\epsilon_s\rangle=20$ for $M_{\rm rms}=0.1$, respectively. Panels $({\rm c})$ and $({\rm d})$ show $\epsilon/\langle\epsilon\rangle=20$ and $\epsilon_d/\langle\epsilon_d\rangle=20$ for $M_{\rm rms}=10$, respectively.}
	\label{fig:isosurface}
\end{figure}
We now use the DNS data at different $M_{\rm rms}$ to investigate the dissipative-anomaly behavior of the total kinetic-energy dissipation rate $\epsilon({\bm x},t)$ and its solenoidal $\epsilon_s({\bm x},t)$ and dilatational $\epsilon_d({\bm x},t)$ components~\citep{Huang:JFM1995}, given by 
\begin{align}
\epsilon({\bm x},t) &= {\bm \sigma}:{\bm \nabla}{\bm u} = 2\mu\, {\bm S}:{\bm S} - \frac{2\mu}{3}({\bm \nabla}\cdot{\bm u})^2 = 2\mu\, {\bm D}:{\bm D}\,, 
\label{eq:total_dissi} \\
\epsilon_s({\bm x},t) &= \mu {\bm \omega}^2\,, 
\label{eq:solenoidal_dissi} \\
\epsilon_d({\bm x},t) &= \frac{4\mu}{3}({\bm \nabla}\cdot{\bm u})^2\,, 
\label{eq:dilatation_dissi}
\end{align}
where ${\bm \omega} = {\bm \nabla}\times{\bm u}$ is the local vorticity. A detailed derivation of this decomposition is given in \citet{Alam_2025,AFS2026}. As shown in these references, a third component is contained in the decomposition, the inhomogeneous dissipation field component $\epsilon_I({\bm x},t)$. The latter drops out when averages are taken. We will not further detail $\epsilon_I$ in the present work. The corresponding solenoidal and dilatational velocity statistics are obtained from the Helmholtz decomposition ${\bm u}={\bm u}_s+{\bm u}_d$ with ${\bm \nabla}\cdot{\bm u}_s=0$ and ${\bm \nabla}\times{\bm u}_d=0$. Here, ${\bm u}_s$ and ${\bm u}_d$ are the solenoidal and dilatational velocity fields, respectively. Figure \ref{fig:isosurface} shows instantaneous isosurface plots of the total, solenoidal and dilatational dissipation fields for the highest Reynolds number. We observe significant structural differences for sub- and supersonic regimes. For $M_{\rm rms}=0.1$, high-amplitude regions of $\epsilon$ manifest as patchy shear layers, while $\epsilon_s$ displays a characteristic tangle of vortex tubes \citep{Zinchenko_2024}. For $M_{\rm rms}=10$, $\epsilon$ and $\epsilon_d$ are almost identical; high-amplitude regions are arranged in the form of two-dimensional curved sheets. We will come back to this point when discussing the geometric analysis in section \ref{sec:multi}.  

For the analysis that follows, the mean total, solenoidal and dilatational dissipation rates are normalized by $\langle\rho\rangle u_{\rm rms}^3/L$, $\langle\rho\rangle u_{s,\rm rms}^3/L_s$ and $\langle\rho\rangle u_{d,\rm rms}^3/L_d$, respectively. Here, $u_{s,\rm rms}=\sqrt{\langle{\bm u}_s^{2}\rangle}$ and $u_{d,\rm rms}=\sqrt{\langle{\bm u}_d^{2}\rangle}$ are the rms velocities of the solenoidal and dilatational fields, and $L_s$ and $L_d$ are their corresponding integral length scales, evaluated in the same manner as $L$ in eq. \eqref{eq:int_length}.

A dissipative anomaly in a turbulent flow exists if 
\begin{equation}
\lim_{Re\to\infty}\frac{\langle \epsilon_\alpha\rangle}{\langle \rho\rangle u^3_{\alpha,{\rm rms}} / L_{\alpha}} \sim Re^{-p} \quad \mbox{for}\quad \alpha = \{\cdot,s,d\}\,.  
\end{equation}
The anomaly is termed {\em strong} dissipative anomaly if $p=0$, i.e.~the normalized mean kinetic energy dissipation rate becomes a constant with increasing Reynolds number; it is termed {\em weak} dissipative anomaly if $0<p<1$. The case of $p=1$ would correspond to the laminar flow, see e.g.~\citet{Eyink2026}.

\begin{figure}
	\begin{center}
	\includegraphics[width=0.9\linewidth]{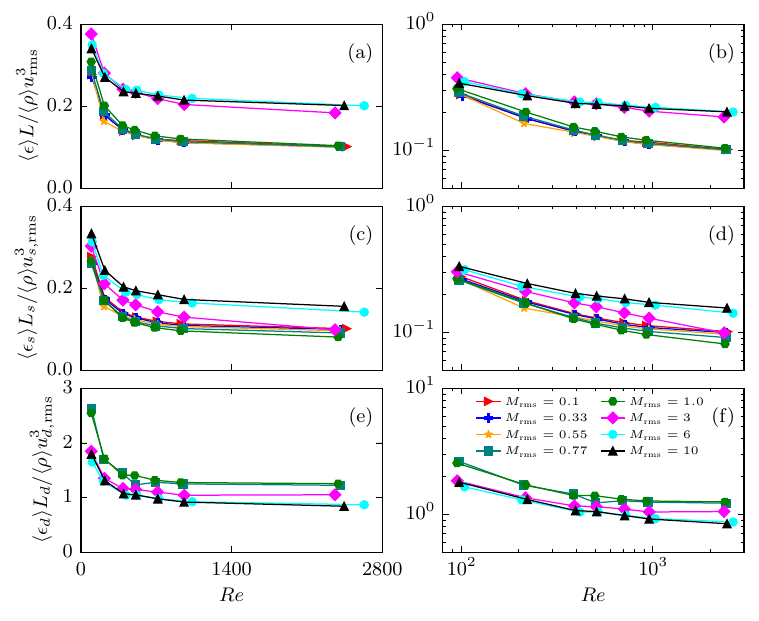}
	\end{center}
	\caption{Total dissipation rate and its components versus Reynolds number for different rms Mach numbers. (a, b) total dissipation rate, (c, d) solenoidal dissipation rate, and (e, f) dilatational dissipation rate. The dissipation rates are normalized using the corresponding integral length scale and rms velocity. The normalized dissipation rates are shown in linear plots in the left column panels (a), (c), and (e) and in double-logarithmic plots in the right column panels (b), (d), and (f).}
	\label{fig:dissi_anomaly}
\end{figure}
\begin{figure}
	\begin{center}
	\includegraphics[width=0.95\linewidth]{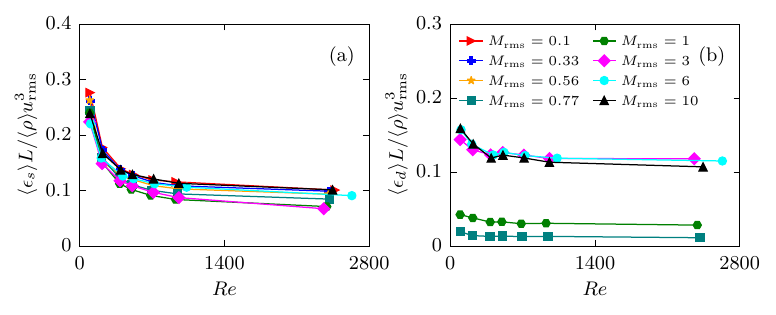}
	\end{center}
	\caption{Solenoidal and dilatational dissipation rates  versus Reynolds number for different rms Mach numbers. (a) solenoidal dissipation rate and (b) dilatational dissipation rate. Here, the dissipation rates are normalized by $\langle \rho \rangle u^3_{\rm rms}/L$.}
	\label{fig:dissi_anomaly_eps_s_eps_d}
\end{figure}

Figure~\ref{fig:dissi_anomaly} shows the normalized total, solenoidal, and dilatational dissipation rates as functions of the Reynolds number $Re$. For the dilatational dissipation field, only cases with $M_{\rm rms} \geq 0.77$ are shown, since dilatational effects are negligible at lower Mach numbers in our solenoidally forced simulations. In the linear representation shown in the left column, all three dissipation rates appear to approach finite asymptotic values as $Re$ increases. The corresponding double-logarithmic representation in the right column reveals, however, that the dissipation rates may still decrease weakly with increasing $Re$. A power-law decay with $Re$ would be indicated by a linear relation. The two distinct branches observed for the total dissipation rate $\epsilon$ remain clearly visible. Overall, the data provide empirical evidence for the existence of a dissipative anomaly in isothermal compressible turbulence. However, because the present simulations reach only moderate Reynolds numbers of $Re\lesssim 2400$, they do not allow us to distinguish conclusively between a weak and a strong anomaly. We mention here that recent DNS of 3D incompressible fluid turbulence by \citet{Iyer2025} indicate a weak dissipative anomaly. To this end, the authors had to cover a wide range up to very high Reynolds numbers.

A particularly striking feature is that the apparent asymptotic state is not unique. Instead, the data separate into two distinct branches. One branch is populated by the subsonic and transonic cases ($M_{\rm rms} \le 1$), whereas the other comprises the highly supersonic cases ($M_{\rm rms} = 6$ and $10$); both approach different constant values at large $Re$. The $M_{\rm rms} =3$ case lies between these two branches, suggesting a gradual transition from the low- to the high- Mach number regime. This is, however, observable for the solenoidal and dilatational components only; for the total dissipation, the $M_{\rm rms} =3$ case is clearly assignable to the supersonic branch. The branch of low-Mach-number is consistent with the subsonic ($M_{\rm rms} \le 0.8$) solenoidally forced simulations of \citet{JDSJFM2021}, which exhibit the same  behavior as incompressible turbulence.

Figure \ref{fig:dissi_anomaly_eps_s_eps_d} rescales solenoidal and dilatational dissipation with integral scale $L$ and total rms velocity $u_{\rm rms}$ to test the sensitivity of the results with respect to the normalization. Differently to figure \ref{fig:dissi_anomaly}, we observe a better collapse of the data of the solenoidal dissipation rate $\epsilon_s$ to one branch, independent of the Mach number. The values of the dilatational dissipation rate $\epsilon_d$ collapse for $M_{\rm rms}\ge 3$ only. This is again different from the results in panels (e) and (f) of figure \ref{fig:dissi_anomaly}. Both figures thus demonstrate how sensitive the results for the Reynolds-number dependence are with respect to the choice of the normalizing quantities.

The existence of two branches with different asymptotic states naturally raises the question of their physical origin. In incompressible turbulence, \citet{sreeni1984,sreeni98b} also obtained different asymptotic values of the normalized dissipation in experiments with different grid configurations \citep{sreeni1984} and in DNS with different forcing \citep{sreeni98b}. He suggested that these differences may arise from the distinct large-scale flow structures produced by the grid configuration or forcing. In the present DNS study, the forcing is identical for all simulations, and the observed branching arises solely from varying the turbulent Mach number. 

We therefore have to examine whether the emergence of the two branches can be traced back to a shift in the dominant nonlinear production mechanisms of the dissipation-rate fields and in the corresponding spatial organization. To this end, we will analyse the nonlinear production terms appearing in the balance equations for the total, solenoidal, and dilatational dissipation-rate fields in the next section. The mean values of these terms will give us a hint to the relative importance of the different production mechanisms, while their spatial distributions identify the flow structures associated with them. Together, these analyses provide a physical interpretation of the two branches.

\section{Nonlinear generation mechanisms of energy dissipation fields}
\label{sec:production}
\subsection{Balances and different production terms}
The balance equations for $\epsilon$, $\epsilon_s$, and $\epsilon_d$ are obtained from Eqs.~\eqref{eq:mass} and \eqref{eq:mom}, together with $\mu = \nu \rho$, note that $\nu$ is constant in our simulations. They are as follows,
\begin{align}
\frac{\partial \epsilon}{\partial t} + {\bm \nabla}\cdot({\bm u}\epsilon) ={}& \mathcal{P}_\epsilon  -4\nu\,{\bm D}:[({\bm \nabla}\otimes{\bm \nabla})p] +\frac{4\nu}{\rho}\, {\bm D}: \left( {\bm \nabla}p\otimes{\bm \nabla}\rho \right) \nonumber\\
& +4\nu\rho\,{\bm D}:{\bm \nabla}\left( \frac{1}{\rho}{\bm \nabla}\cdot{\bm \sigma} \right) +4\nu\rho\,{\bm D}:{\bm \nabla}{\bm f}\,,
\label{eq:eps}
\\[2mm]
\frac{\partial \epsilon_s}{\partial t} +{\bm \nabla}\cdot({\bm u}\epsilon_s) ={}& \mathcal{P}_{\epsilon_s} +\frac{2\nu}{\rho}\, {\bm \omega}\cdot \left( {\bm \nabla}\rho\times{\bm \nabla}p \right)
+2\nu\rho\,{\bm \omega}\cdot \left[ {\bm \nabla}\times \left( \frac{1}{\rho}{\bm \nabla}\cdot{\bm \sigma} \right) \right] \nonumber\\
&+2\nu\rho\,{\bm \omega}\cdot \left( {\bm \nabla}\times{\bm f} \right)\,,
\label{eq:eps_s}
\\[2mm]
\frac{\partial \epsilon_d}{\partial t} +{\bm \nabla}\cdot({\bm u}\epsilon_d) ={}& \mathcal{P}_{\epsilon_d}
- \frac{8}{3}\nu\, ({\bm \nabla}\cdot{\bm u}) \biggl[{\bm \nabla}^{2}p -  \frac{1}{\rho} {\bm \nabla}\rho\cdot{\bm \nabla}p \biggr] \nonumber\\
& + \frac{8}{3}\nu\rho\, ({\bm \nabla}\cdot{\bm u}) \biggl[ {\bm \nabla}\cdot \left( \frac{1}{\rho}{\bm \nabla}\cdot{\bm \sigma} \right)  + {\bm \nabla}\cdot{\bm f} \biggr]\,,
\label{eq:eps_d}
\end{align}
where the nonlinear production terms $\mathcal{P}_\epsilon$, $\mathcal{P}_{\epsilon_s}$, and $\mathcal{P}_{\epsilon_d}$ are
\begin{align}
\mathcal{P}_\epsilon={}& -4\nu\rho\,{\rm tr}({\bm D}^3) +\nu\rho\,({\bm \omega}\otimes{\bm \omega}):{\bm D} -\frac{8}{3}\nu\rho\, ({\bm \nabla}\cdot{\bm u})\,{\bm D}:{\bm D}\,,
\label{eq:p_eps}
\\[2mm]
\mathcal{P}_{\epsilon_s} ={}& 2\nu\rho\,({\bm \omega}\otimes{\bm \omega}):{\bm D} -\frac{4}{3}\nu\rho\,{\bm \omega}^2 ({\bm \nabla}\cdot{\bm u})\,,
\label{eq:p_eps_s}
\\[2mm]
\mathcal{P}_{\epsilon_d} ={}& - \frac{8}{3}\nu\rho\, ({\bm \nabla}\cdot{\bm u}) {\bm D}:{\bm D} +\frac{4}{3} \nu\rho {\bm \omega}^2 ({\bm \nabla}\cdot{\bm u}) -\frac{8}{9} \nu\rho ({\bm \nabla}\cdot{\bm u})^{3}\,.
\label{eq:p_eps_d}
\end{align}
Here, ${\rm tr}(\cdot)$ denotes the trace operator. These production terms arise from five distinct nonlinear mechanisms: (i) deviatoric strain-rate self-interaction, ${\rm tr}({\bm D}^3)$; (ii) vortex stretching and tilting, $({\bm \omega}\otimes{\bm \omega}):{\bm D}$; (iii) strain--dilatation interaction, $({\bm \nabla}\cdot{\bm u})\,{\bm D}:{\bm D}$; (iv) vorticity--dilatation interaction, ${\bm \omega}^2 ({\bm \nabla}\cdot{\bm u})$; and (v) pure dilatation, $({\bm \nabla}\cdot{\bm u})^{3}$. Among these, $\mathcal{P}_{\epsilon}$ comprises contributions from the first three mechanisms, $\mathcal{P}_{\epsilon_s}$ from the second and fourth mechanisms, and $\mathcal{P}_{\epsilon_d}$ from the last three mechanisms. The remaining terms in Eqs.~\eqref{eq:eps}-\eqref{eq:eps_d} arise from pressure gradients, density gradients, viscous stresses, and external forcing. To distinguish the contributions arising from the solenoidal and dilatational motions, we further decompose the nonlinear production terms using the Helmholtz decomposition of the velocity field, ${\bm u} = {\bm u}_s + {\bm u}_d$, which yields the corresponding decomposition of the deviatoric strain-rate tensor, ${\bm D} = {\bm D}_s + {\bm D}_d$ where
\begin{align}
{\bm D}_s = \frac{1}{2}[{\bm \nabla} {\bm u}_s + ( {\bm \nabla} {\bm u}_s)^{T}] \quad {\rm and} \quad {\bm D}_d =  \frac{1}{2}[{\bm \nabla} {\bm u}_d + ( {\bm \nabla} {\bm u}_d)^{T}] - \frac{1}{3} ({\bm \nabla} \cdot {\bm u}_d) {\bm I}\,.
\end{align}
Substituting this decomposition into Eqs.~\eqref{eq:p_eps}--\eqref{eq:p_eps_d} yields
\begin{align}
\mathcal{P}_{\epsilon}={}& \mathcal{P}_{\epsilon,1} + \mathcal{P}_{\epsilon,2} + \mathcal{P}_{\epsilon,3} + \mathcal{P}_{\epsilon,4} + \mathcal{P}_{\epsilon,5} + \mathcal{P}_{\epsilon,6} + \mathcal{P}_{\epsilon,7} + \mathcal{P}_{\epsilon,8} + \mathcal{P}_{\epsilon,9}\,,
\label{eq:p_eps_comp}
\\[2mm]
\mathcal{P}_{\epsilon_s} ={}& \mathcal{P}_{\epsilon_s,1} + \mathcal{P}_{\epsilon_s,2} + \mathcal{P}_{\epsilon_s,3}\,,
\label{eq:p_eps_s_comp}
\\[2mm]
\mathcal{P}_{\epsilon_d} ={}& \mathcal{P}_{\epsilon_d,1} + \mathcal{P}_{\epsilon_d,2} + \mathcal{P}_{\epsilon_d,3} + \mathcal{P}_{\epsilon_d,4} + \mathcal{P}_{\epsilon_d,5}\,,
\label{eq:p_eps_d_comp}
\end{align}
where
\begin{equation}
\left.
\begin{aligned}
\mathcal{P}_{\epsilon,1}=& -4\nu\rho\,{\rm tr}({\bm D}_s^3)\,, \quad
\mathcal{P}_{\epsilon,2}= -12\nu\rho\,{\rm tr}({\bm D}_s^2{\bm D}_d)\,, \\
\mathcal{P}_{\epsilon,3}=& -12\nu\rho\,{\rm tr}({\bm D}_s{\bm D}_d^2)\,, \quad
\mathcal{P}_{\epsilon,4}= -4\nu\rho\,{\rm tr}({\bm D}_d^3)\,, \\
\mathcal{P}_{\epsilon,5}=& \ \frac{\mathcal{P}_{\epsilon_s,1}}{2} = \nu\rho\,({\bm \omega}\otimes{\bm \omega}):{\bm D}_s\,, \\
\mathcal{P}_{\epsilon,6}= & \ \frac{\mathcal{P}_{\epsilon_s,2}}{2}= \nu\rho\,({\bm \omega}\otimes{\bm \omega}):{\bm D}_d\,, \\
\mathcal{P}_{\epsilon,7}=& \ \mathcal{P}_{\epsilon_d,1} = -\frac{8}{3}\nu\rho\, ({\bm \nabla}\cdot{\bm u})\,{\bm D}_s:{\bm D}_s\,, \\
\mathcal{P}_{\epsilon,8}= & \ \mathcal{P}_{\epsilon_d,2} = -\frac{8}{3}\nu\rho\, ({\bm \nabla}\cdot{\bm u})\,{\bm D}_s:{\bm D}_d\,, \\
\mathcal{P}_{\epsilon,9}=& \ \mathcal{P}_{\epsilon_d,3} = -\frac{8}{3}\nu\rho\, ({\bm \nabla}\cdot{\bm u})\,{\bm D}_d:{\bm D}_d\,,\\
\mathcal{P}_{\epsilon_s,3}=& -\mathcal{P}_{\epsilon_d,4}= -\frac{4}{3}\nu\rho\,{\bm \omega}^2 ({\bm \nabla}\cdot{\bm u})\,, \\
\mathcal{P}_{\epsilon_d,5}=& -\frac{8}{9}\nu\rho ({\bm \nabla}\cdot{\bm u})^{3}\,.
\end{aligned}
\right\}
\label{eq:p_comp_def}
\end{equation}
This decomposition separates the production into contributions arising from purely solenoidal motions, purely dilatational motions, and their interactions. Purely solenoidal contributions involve only the solenoidal fields ${\bm D}_s$ and ${\bm \omega}$, whereas purely dilatational contributions involve only the dilatational fields ${\bm D}_d$ and ${\bm \nabla} \cdot {\bm u}$. Contributions arising from interactions between these fields are referred to as mixed contributions.

\begin{figure}
	\begin{center}
		\includegraphics{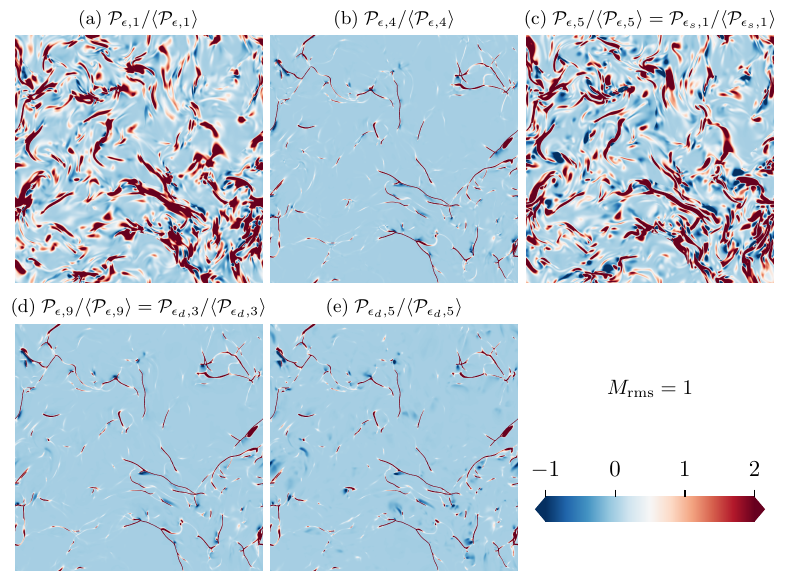}
	\end{center} 
	\caption{At $M_{\rm rms}=1$, snapshot contour slices of the production terms resulting from solenoidal motions (a, c) and from dilatational motions (b, d, e), plotted in units of their respective mean values. The mixed terms are not shown; their visualizations exhibit a combination of the incompressible-like structures seen in panels (a, c) and the shocklet structures observed in panels (b, d, e). The definitions of all production terms are provided in \eqref{eq:p_comp_def}. The Reynolds number of the three-dimensional flow is $Re=2400$.}
	\label{fig:contour_production_terms_M_1}
\end{figure}
\begin{figure}
	\begin{center}
		\includegraphics{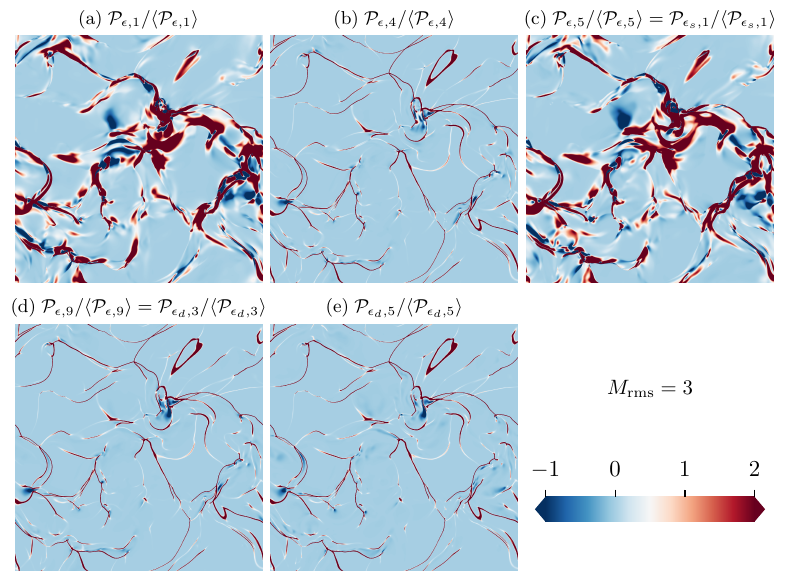}
	\end{center}
	\caption{Same as figure~\ref{fig:contour_production_terms_M_1}, but for $M_{\rm rms}=3$.}
	\label{fig:contour_production_terms_M_3}
\end{figure}
\begin{figure}
	\begin{center}
		\includegraphics{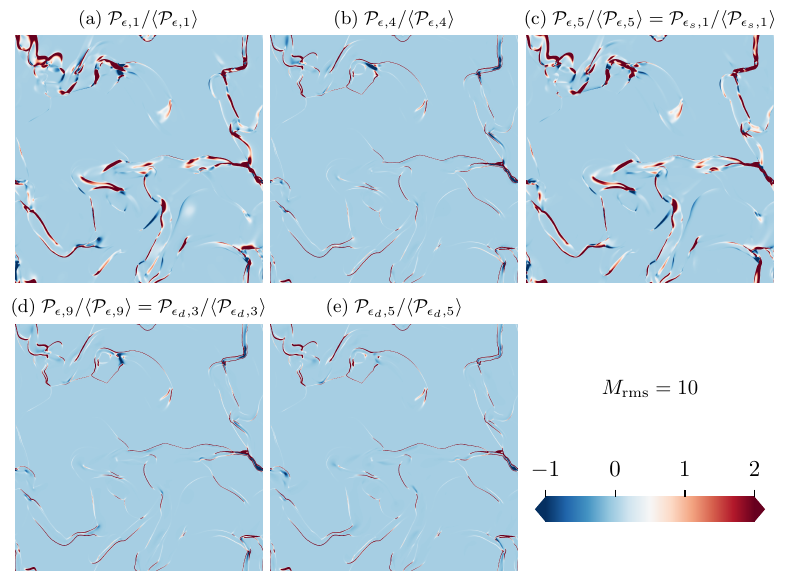}
	\end{center}
	\caption{Same as figure~\ref{fig:contour_production_terms_M_1}, but for $M_{\rm rms}=10$.}
	\label{fig:contour_production_terms_M_10}
\end{figure}
Figures \ref{fig:contour_production_terms_M_1}, \ref{fig:contour_production_terms_M_3} and \ref{fig:contour_production_terms_M_10} show two-dimensional contour slice plots of the decomposed solenoidal and dilatational production terms at $M_{\rm rms}=1$, $3$ and $10$, respectively. The data are obtained for the highest Reynolds number in our series, $Re=2400$.  At $M_{\rm rms}=1$, the production associated with the pure solenoidal mechanisms of solenoidal strain-rate interaction ${\rm tr}({\bm D}_s^3)$ in panel (a) and solenoidal vortex stretching $({\bm \omega}\otimes{\bm \omega}):{\bm D}_s$ in panel (c) exhibits the shear-layer and vortical structures characteristic of incompressible turbulence. In contrast, the various pure dilatational mechanisms, arising from the dilatational strain-rate interaction ${\rm tr}({\bm D}_d^3)$ in panel (b), the interaction between the dilatational strain rate and dilatation $({\bm \nabla} \cdot {\bm u}) {\bm D}_d:{\bm D}_d$ in panel (d), and pure dilatation $(\bm \nabla \cdot {\bm u})^3$ in panel (e), exhibit nearly identical structures dominated by fragmented shocklets distributed throughout the flow. As $M_{\rm rms}$ increases further into the supersonic regime, the spatial organization of these structures changes substantially. At $M_{\rm rms}=3$ in figure~\ref{fig:contour_production_terms_M_3}, the fragmented shocklets are replaced by elongated shock layers, around which the solenoidal production also becomes localized. The incompressible-like structures in panels (a) and (c) are still present, but are largely confined to the vicinity of the shock layers and thus occupy a much smaller fraction of the flow domain than up to $M_{\rm rms}=1$. At $M_{\rm rms}=10$ in figure~\ref{fig:contour_production_terms_M_10}, all production terms become strongly localized within a small number of dominant shock layers, with negligible production elsewhere. The incompressible-like  structures are no longer evident, and the solenoidal and dilatational production exhibit remarkably similar spatial distributions.

Taken together, these visualizations suggest that $M_{\rm rms}=3$ lies within a structural transition from the incompressible-like flow, in which fragmented shocklets emerge at $M_{\rm rms}\sim1$, to the strongly shock-layer-dominated regime established at higher Mach numbers, where the production is concentrated almost entirely within a few dominant shock layers. We now examine the mean contributions of the individual production mechanisms across different $M_{\rm rms}$. Since the characteristic flow scales vary with Mach number, an appropriate nondimensionalization is required. We therefore nondimensionalize \eqref{eq:eps} using
\begin{equation}
t^* = \frac{t}{L/u_{\rm rms}}, \quad
{\bm x}^* = \frac{{\bm x}}{L}, \quad
{\bm u}^* = \frac{{\bm u}}{u_{\rm rms}}, \quad
\epsilon^* = \frac{\epsilon}{\langle \rho\rangle u_{\rm rms}^3/L}.
\end{equation}
Substituting these variables into the left-hand side of \eqref{eq:eps} gives
\begin{equation}
\frac{\partial \epsilon}{\partial t}
+ {\bm \nabla}\cdot({\bm u}\epsilon)
=
\frac{\langle \rho\rangle u_{\rm rms}^4}{L^2}
\left[
\frac{\partial \epsilon^*}{\partial t^*}
+ {\bm \nabla}^*\cdot({\bm u}^*\epsilon^*)
\right].
\end{equation}
Dividing \eqref{eq:eps} by $\langle \rho \rangle u_{\rm rms}^4/L^2$ yields the nondimensional production term,
\begin{equation}
\mathcal{P}_{\epsilon}^{*} = \frac{\mathcal{P}_{\epsilon}}{\langle \rho \rangle u_{\rm rms}^{4}/L^{2}} = \sum_{i=1}^{9}\mathcal{P}_{\epsilon,i}^{*}\,, \qquad \mathcal{P}_{\epsilon,i}^{*} = \frac{\mathcal{P}_{\epsilon,i}} {\langle \rho \rangle u_{\rm rms}^{4}/L^{2}}\,.
\label{eq:p_eps_nondim}
\end{equation}
The solenoidal and dilatational production terms are nondimensionalized in an analogous manner using the characteristic solenoidal and dilatational velocity and length scales, respectively,
\begin{align}
\mathcal{P}^*_{\epsilon_s} &= \frac{\mathcal{P}_{\epsilon_s}} {\langle \rho \rangle u_{s,\rm rms}^4/L_s^2} = \sum_{i=1} ^{3}\mathcal{P}^*_{\epsilon_s,i}\,, \qquad \mathcal{P}^*_{\epsilon_s,i} = \frac{\mathcal{P}_{\epsilon_s,i}} {\langle \rho \rangle u_{s,\rm rms}^4/L_s^2}\,,
\label{eq:p_eps_s_nondim} \\
\mathcal{P}^*_{\epsilon_d} &= \frac{\mathcal{P}_{\epsilon_d}} {\langle \rho \rangle u_{d,\rm rms}^4/L_d^2} = \sum_{i=1} ^{5}\mathcal{P}^*_{\epsilon_d,i}, \qquad \mathcal{P}^*_{\epsilon_d,i} = \frac{\mathcal{P}_{\epsilon_d,i}} {\langle \rho \rangle u_{d,\rm rms}^4/L_d^2}.
\label{eq:p_eps_d_nondim}
\end{align}

\begin{figure}
	\begin{center}
		\includegraphics{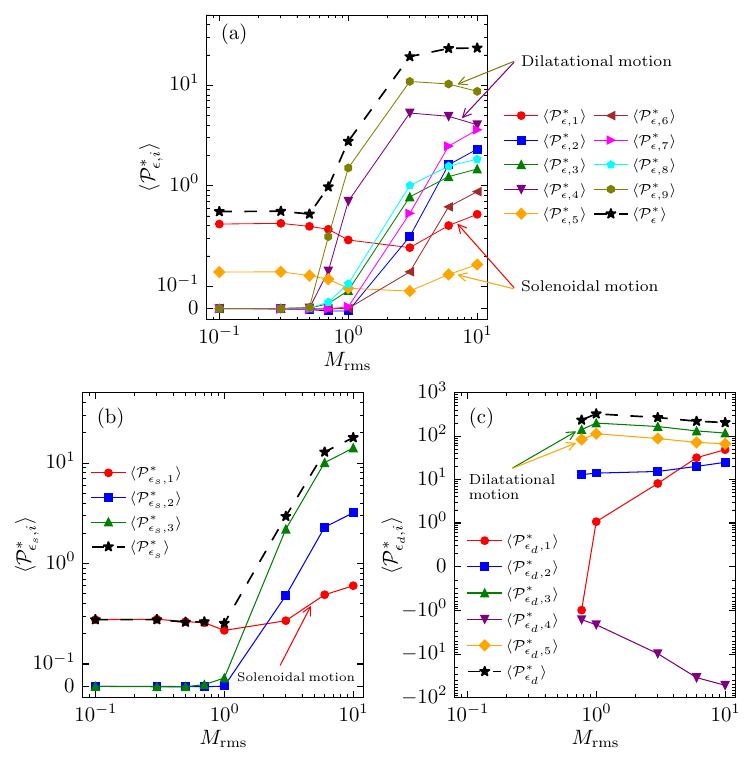}
	\end{center}
	\caption{Mean nondimensional nonlinear production terms as functions of the rms Mach number. (a) Contributions of the different components to the production of the total dissipation rate, $\langle \mathcal{P}_{\epsilon,i}^{*} \rangle$, together with the total production, $\langle \mathcal{P}_{\epsilon}^{*} \rangle$. (b) Contributions of the different components to the production of the solenoidal dissipation rate, $\langle \mathcal{P}_{\epsilon_s,i}^{*} \rangle$, together with the total production, $\langle \mathcal{P}_{\epsilon_s}^{*} \rangle$. (c) Contributions of the different components to the production of the dilatational dissipation rate, $\langle \mathcal{P}_{\epsilon_d,i}^{*} \rangle$, together with the total production, $\langle \mathcal{P}_{\epsilon_d}^{*} \rangle$. The arrows indicate the production terms arising purely from solenoidal and purely from dilatational motions. The remaining terms, not indicated by the arrows, correspond to mixed contributions arising through different interaction mechanisms between solenoidal and dilatational motions. See \eqref{eq:p_eps_nondim}--\eqref{eq:p_eps_d_nondim}, together with \eqref{eq:p_comp_def}, for the definitions of the terms. Data are again for $Re=2400$.}
\label{fig:production_terms}
\end{figure}
Figure~\ref{fig:production_terms} summarizes the mean nondimensional production terms, which, together with the spatial organization of the production mechanisms, help interpret the two branches in figure~\ref{fig:dissi_anomaly}. For $M_{\mathrm{rms}}\leq0.55$, the production of the total dissipation in panel (a) is almost entirely due to the purely solenoidal terms, while the purely dilatational and mixed contributions are negligible. This reflects the nearly incompressible nature of the flow under purely solenoidal forcing. At $M_{\mathrm{rms}}=0.77$, the purely dilatational production becomes appreciable and, by $M_{\mathrm{rms}}=1$, it already exceeds the purely solenoidal production. Nevertheless, the $M_{\mathrm{rms}}=1$ case still belongs to the low-Mach-number branch in figure~\ref{fig:dissi_anomaly}. This indicates that a larger mean dilatational contribution alone is insufficient to account for the different branches of the normalized dissipation. At $M_{\mathrm{rms}}=1$, as seen in figure~\ref{fig:contour_production_terms_M_1}, the dilatational production is concentrated in fragmented shocklets distributed throughout an otherwise dominantly vortical flow. By contrast, in the highly supersonic cases, all production fields become concentrated within a few dominant shock layers, including the purely solenoidal contribution, as seen in figure~\ref{fig:contour_production_terms_M_10}. This structural reorganization is accompanied by a change in the mean production budget: the purely dilatational contribution becomes substantially dominant, and the mixed contributions also exceed the purely solenoidal term. These observations suggest that the transition between the two asymptotic branches requires not only a change in the relative magnitude of the production mechanisms but also a reorganization of the production into a few dominant shock layers. This result is consistent with the interpretation of \citet{sreeni98b} that different asymptotic states of the normalized dissipation are associated with differences in the underlying flow structure.

The production of $\epsilon_s$ contains one purely solenoidal contribution, $P_{\epsilon_s,1}$, associated with solenoidal vortex stretching, and two mixed contributions, $P_{\epsilon_s,2}$ and $P_{\epsilon_s,3}$, associated with dilatational vortex stretching and the vorticity--dilatation interaction, respectively. As shown in figure~\ref{fig:production_terms}(b), the production for $M_{\rm rms}\le1$ is governed almost entirely by the solenoidal vortex-stretching term, while both mixed contributions remain negligible. The spatial distribution of this dominant contribution retains the familiar incompressible-like vortical structures shown in figure~\ref{fig:contour_production_terms_M_1}(c). 

These results show that solenoidal dynamics remain close to those of incompressible flow up to $M_{\rm rms}=1$. This behavior explains why the solenoidal dissipative anomaly remains on the low-Mach-number asymptotic branch over this Mach-number range, as shown in figure~\ref{fig:dissi_anomaly}(c). A different picture emerges in the highly supersonic regime ($M_{\rm rms}=6$ and $10$). In this regime, the production fields become concentrated almost entirely within a few dominant shock layers, as shown in figure~\ref{fig:contour_production_terms_M_10}. The dominant contribution to the production also arises from the coupling of the solenoidal and dilatational motions, with the vorticity--dilatation interaction providing the largest contribution. The emergence of this dilatation-dominated shock-layer state gives rise to the high-Mach-number branch of the solenoidal dissipative anomaly shown in figure~\ref{fig:dissi_anomaly}(c). The $M_{\rm rms}=3$ case is characterized by the coexistence of incompressible-like vortical structures and elongated shock layers, despite the dominance of the mixed production terms. This mixed structural organization gives rise to the intermediate asymptotic state shown in figure~\ref{fig:dissi_anomaly}(c).

Figure~\ref{fig:production_terms}(c) presents the production terms for $\epsilon_d$ from $M_{\rm rms}=0.77$ onwards, where dilatational effects become appreciable. The production of $\epsilon_d$ contains two purely dilatational contributions, $P_{\epsilon_d,3}$ and $P_{\epsilon_d,5}$, while the remaining three terms arise from the interaction between the solenoidal and dilatational motions. We observe a strong increase of $\langle \mathcal{P}_{\epsilon_d,1}^{*} \rangle$ with Mach number, which remains smaller compared to the two former terms. As shown in the figure, the production is dominated by the purely dilatational contributions, with the mixed contributions also increasing in the highly supersonic regime. Nevertheless, the dilatational dissipative anomaly also exhibits two branches, as shown in figure~\ref{fig:dissi_anomaly}(e). However, unlike $\epsilon_s$, whose low-Mach-number branch is associated with incompressible-like vortical structures, the low-Mach-number branch of $\epsilon_d$ is characterized by fragmented shocklets. At $M_{\rm rms}=3$, the production is distributed over a dense network of elongated shock layers. By contrast, at $M_{\rm rms}=10$, it is concentrated in only a few dominant shock layers, leaving large regions of the domain with negligible production, as shown in figures~\ref{fig:contour_production_terms_M_3} and \ref{fig:contour_production_terms_M_10}. This change in the shock-layer organization is associated with the transition from the low-Mach-number branch through the intermediate state at $M_{\rm rms}=3$ to the high-Mach-number branch.

\subsection{Where do we stand so far?} 
Let us summarize and reflect our preliminary findings here. We started with the Reynolds-number dependence of the normalized mean dissipation rate and its components. In incompressible turbulence, the empirical signature of a dissipative anomaly is that $ \langle \epsilon \rangle L/u_{\rm rms}^3$ approaches a finite value at sufficiently large Reynolds numbers \citep{sreeni1984,sreeni98b}. However, in compressible turbulence, this seems to be less strict. The dissipation field contains both solenoidal and dilatational contributions, and compressive motions introduce additional density- and pressure-related ways for energy transfer. This certainly also holds for the simplest realization, the isothermal case that is considered here.

Figure \ref{fig:dissi_anomaly} showed the normalized total dissipation rate, together with its solenoidal and dilatational components, as functions of Reynolds number. The data did not collapse onto a single line across all Mach numbers. Instead, they separated into two distinct branches. For the total and solenoidal dissipation rates, the subsonic and transonic cases form the lower branch, whereas the highly supersonic cases form the upper branch. This ordering is reversed for the dilatational dissipation rate, for which the subsonic and transonic cases lie on the upper branch. These branches distinguish the predominantly incompressible-like organization at lower Mach numbers from the shock-layer-dominated organization in the highly supersonic regime.

This result can be compared to the data collected by \citet{JDSJFM2021}. Their simulations covered a Mach-number range $0.05\le M_{\rm rms}\le 0.8$, which corresponds to the subsonic flow regime, where vortex structures are dominant. The data thus could not detect the second high-Mach-number branch observed in figure \ref{fig:dissi_anomaly}. For solenoidally-forced turbulence, they found a trend similar to incompressible turbulence. We confirm this result here. Larger deviations appeared when the flow was driven by dilatational forcing. In the previous subsection, we disentangled the various terms that generate energy dissipation, showing that the generation mechanisms of dissipation differ for subsonic/transonic and supersonic cases.  

To conclude, all simulations in the present study are driven by purely solenoidal forcing. At low Mach numbers, the normalized dissipation follows a single branch that is close to the incompressible case and is consistent with the solenoidally-forced cases of \cite{JDSJFM2021}. However, a second branch appears as the Mach number increases into the supersonic flow regime ($M>1$). Since the forcing remains unchanged, this separation cannot be explained by different forcings. Instead, it shows an increasing influence of compressive motions on the dissipation. To better understand the two branches observed for the total dissipation, we will analyse the anomalous dissipation terms, cf.~eqns.~\eqref{eq:diss_terms}, in the next subsection, and quantify their magnitude in the vicinity of characteristic flow structures at different Mach numbers.

\subsection{Anomalous dissipation analysis: DNS data}
\begin{figure}
    \begin{center}
        \includegraphics[width=0.95\linewidth]{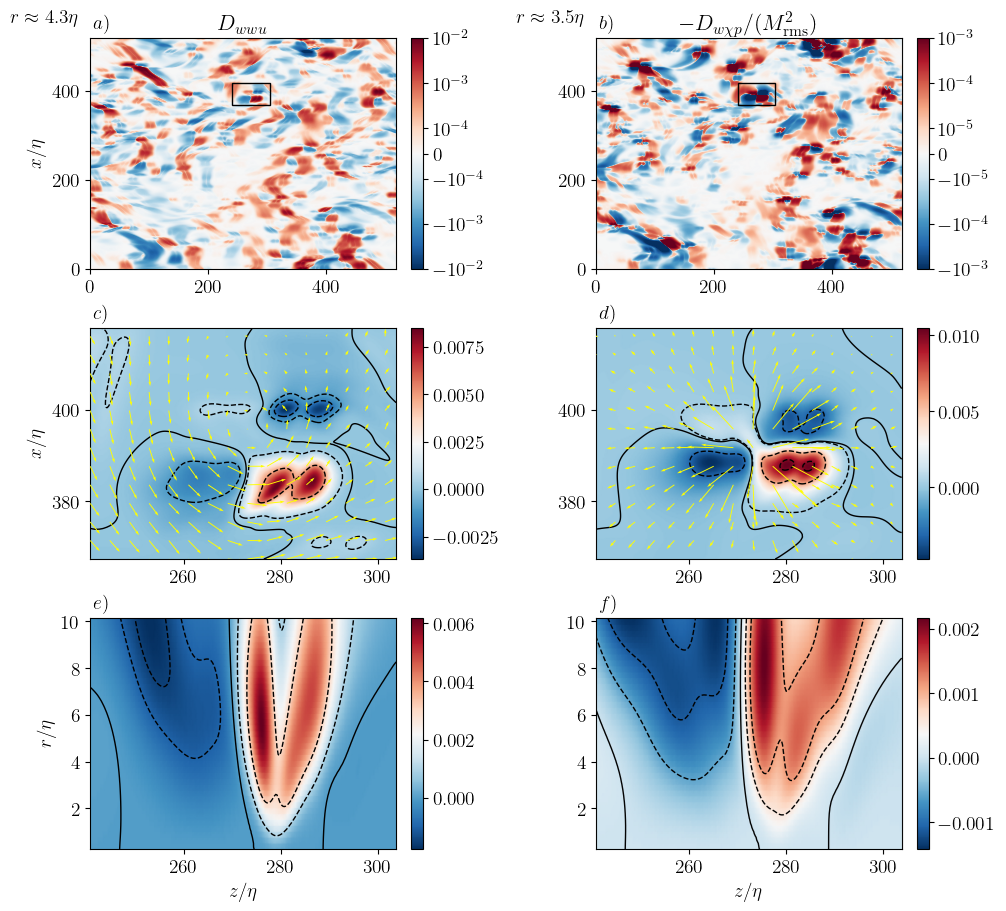}
    \end{center}
    \caption{Local anomalous dissipation distribution for $M_{\rm rms}=0.1$. Panels (a,b) show the corresponding fields in the full $x$-$z$ plane. The black boxes mark the region shown in panels (c,d), which contain one single local high-vorticity event. The yellow arrows in panels (c,d) show the velocity and density gradient vector fields, respectively. Panels (e,f) display the dissipation contributions in the $r$-$z$ plane at $x/\eta\approx 380$. For panels (a)-(d), the coarse-graining scale $r$ is chosen separately at few Kolmogorov lengths. Solid and dashed black lines denote the zero- and nonzero contour levels.}
    \label{fig:local_dissiaption_M_0_1}
\end{figure}

Before starting, we would like to note that the anomalous dissipation terms actually exist in the large-Reynolds-number limit only, which is certainly not obtained at $Re=2400$. We term the fields $D_{wwu}(r,{\bm x})$ and $D_{w\chi p}(r,{\bm x})$ anomalous dissipation contributions, but they cannot be more than {\em precursors} to anomalous dissipation. To calculate dissipation terms in \eqref{eq:DR_balance}, we choose the Gaussian-type test function, see also \citet{Dubrulle2019}. Specifically, we use the first-order normalized Gaussian, given by
\begin{equation}
\label{eq:test_function}
    \psi_1(\xi) = \frac{1}{\sqrt{2\pi}}\frac{\partial}{\partial \xi} [\exp\left(-\xi^2/2\right)] = -\frac{\xi}{\sqrt{2\pi}} \exp(-\xi^2/2).
\end{equation}
In all computations, we take the increment vector only with respect to one direction, the vertical one with $\bm\xi=(0,0,\xi_z)$ and use only the longitudinal velocity, density and pressure increments evaluated over the same vertical separation. The corresponding increments entering $D_{wwu}$ and $D_{w\chi p}$ are calculated from the field values at $(x,y,z+\xi_z)$ and $(x,y,z)$. For the numerical calculation, the local dissipation terms are evaluated over a finite interval of $[-8r,8r]$ in each coordinate direction. The chosen interval contains the effective support of the Gaussian test function, such that the cutoff error of the numerical integration is below $0.1\%$. Using this procedure, we calculate the anomalous-dissipation terms \eqref{eq:diss_deform} and \eqref{eq:diss_barop}, as shown in figure~\ref{fig:local_dissiaption_M_0_1}.

The DNS case $M_{\rm rms}=0.1$ and $Re=2400$ is practically incompressible. The local event in figure \ref{fig:local_dissiaption_M_0_1}(c,d) shows the same vortex-stretching structure as the high-vorticity events that were previously analyzed in detail for incompressible homogeneous isotropic turbulence by \citet{Zinchenko_2024}. In particular, the velocity field forms a localized rotational structure, with the strongest dissipation concentrated around the vortex core, approximately at a distance of the Burgers radius ($\sim \eta$). These low-Mach-number structures can be interpreted as Burgers-vortex events.

\begin{figure}
    \begin{center}
        \includegraphics[width=0.95\linewidth]{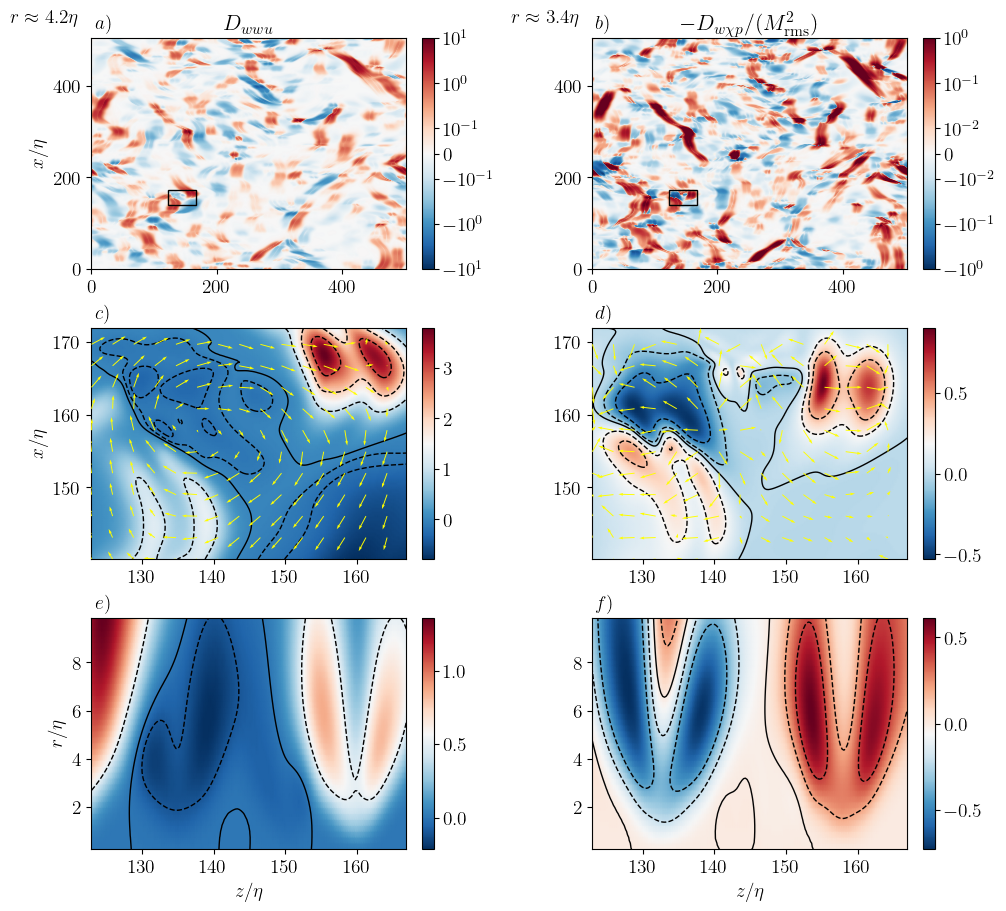}
    \end{center}
    \caption{Local anomalous dissipation distribution for $M_{\rm rms}=1$, compare with figure \ref{fig:local_dissiaption_M_0_1}. Panels (a,b) show the corresponding fields in the full $x$-$z$ plane. The black boxes mark the region shown in panels (c,d), which contains one single local high-vorticity event. Similar to figure \ref{fig:local_dissiaption_M_0_1}, the yellow arrows in panels (c,d) show the velocity and density gradient vector fields respectively. Panels (e,f) show the dissipation contributions in the $r$-$z$ plane at $x/\eta\approx160$. For panels (a)-(d), the coarse-graining scale $r$ is chosen separately at few Kolmogorov lengths. Solid and dashed black lines denote the zero- and nonzero contour levels.}
    \label{fig:local_dissiaption_M_1}
\end{figure}

Next, we perform the same calculation for $M_{\rm rms}=1$ and $Re=2400$. The results can be seen in figure \ref{fig:local_dissiaption_M_1}.
Despite the fully compressible flow, the spatial organization of the anomalous dissipation fields across the full plane remains qualitatively similar to that observed for $M_{\rm rms}=0.1$. Both terms are concentrated in localized structures, and strong events are now associated not only with vortex-like structures, but with shock waves as well. However, the event shown in figure \ref{fig:local_dissiaption_M_1}(c,d) also demonstrates that vortex stretching still contributes significantly to the local energy transfer at $M_{\rm rms}=1$.

Compressibility introduces additional flow structures, which, in the transition to supersonic flows, include shock events. This provides additional channels for energy transfer across scales. Nevertheless, figure~\ref{fig:local_dissiaption_M_1} shows that vortex-related events remain important in the transonic regime. The persistence of vortex events up to transonic Mach numbers may also explain why the normalized total dissipation for $M_{\rm rms}\leq1$ remains on the low-Mach-number branch in figure~\ref{fig:dissi_anomaly}. Within this interpretation, the transition to the second branch at higher Mach numbers is associated with the growing importance of compressible and shock-related dissipation mechanisms. 

The highly supersonic case at $M_{\rm rms}=10$, where shock-layer structures dominate the local dissipation field, is discussed in detail in the next subsection in connection with an analytical Burgers-type shock-layer model. This model will be used since the statistical analysis of moments of the energy dissipation rate revealed in \citet{AFS2026} that the algebraic scaling with respect to the Reynolds number $Re$ is bounded from above by Burgers turbulence \citep{Friedrich2018}. 

\subsection{Anomalous dissipation analysis: Randomly oriented Burgers-type shock layer}
\subsubsection{One-dimensional Burgers-type shock}
The results above suggest that the transition to the high-Mach-number dissipation branch is connected with an increasing contribution from strongly compressible structures. To isolate the anomalous dissipation associated with such structures, we first consider a simplified model of a viscous shock layer in this subsection as a preparation for the subsequent parts. As a simple one-dimensional analogue for the three-dimensional shock-layer events, we consider the viscous Burgers equation in the following, which is given by
\begin{equation}
\partial_t u + u\,\partial_x u = \nu\,\partial_x^2 u.
\label{eq:burgers_eq}
\end{equation}
Its classical traveling-shock solution connecting the constant states $u^+$ and $u^-$ with $u^+>u^-$ is given by
\begin{equation}
u(x,t)=\bar u-\Delta u\tanh\left(\frac{x}{\delta_u}\right),\quad
\bar u=\frac{u^++u^-}{2},\quad
\Delta u=\frac{u^+-u^-}{2},\quad
\delta_u=\frac{2\nu}{\Delta u},
\label{eq:burgers_shock}
\end{equation}
where $\delta_u$ denotes shock-layer thickness. For the inverse square-root density field $\chi=1/\sqrt{\rho}$, we consider a similar profile:
\begin{equation}
    \chi (x,t)=\frac{1}{\sqrt{\rho}}=\bar \chi-\Delta\chi \tanh{\left(\frac{x}{\delta_\rho}\right)}.
\end{equation}
The pressure is then obtained from the isothermal equation of state,
\begin{equation}
    p(x)=c_s^2\rho(x)=\frac{c_s^2}{\chi^2(x)}.
\end{equation}
To calculate dissipation terms in \eqref{eq:DR_balance}, we use again \eqref{eq:test_function}. Now, we can verify that the anomalous dissipation terms vanish with $r^2$ as $r \to 0$ in the viscous subrange for $r<\eta$. To this end, we use \eqref{eq:diss_deform}:
\begin{equation}
    D_{wwu}(r, {x}) = \frac{1}{4r^2} \int_{-\infty}^{+\infty} \psi_1 \left( \frac{\xi}{r} \right)  \delta_\xi w^2 \delta_\xi u \,d \xi
\end{equation}
Introducing $\tilde \xi=\xi/r$, expanding the
increments for small $r$, and setting $\delta_u=\delta_\rho=1$, we obtain
\begin{equation}
    D_{wwu}(r, {x}) = \frac{r^2}{4} \int_{-\infty}^{+\infty} \tilde \xi^3 \psi_1(\tilde \xi)  w'^2(x)  u'(x)d \tilde \xi.
\end{equation}
where the primes denote first spatial derivatives evaluated at $x$, arising from the Taylor expansion. As a result, one gets
\begin{equation}
    \frac{D_{wwu}}{\rho_c\Delta u^3}(r, {x})=\frac{3}{4}\left(\frac{\bar u}{\Delta u}-\frac{\bar\chi}{\Delta\chi}\right)^2\frac{\rho^2(x)}{\rho_c^2\cosh^4{x}}r^2,\quad r\ll 1,\quad\rho_c=\frac{1}{\Delta\chi^2}.
\end{equation}
If we look at a different limit $r\rightarrow\infty$ with the same rescaling $\tilde \xi=\xi/r$, the increments approach their asymptotic jump values. At the shock location, we have
\begin{equation}
    \delta_\xi w^2\delta_\xi u=-\text{sgn}(\tilde \xi)\Delta w^2\Delta u\,.
\end{equation}
Therefore, we obtain
\begin{equation}
    D_{wwu}(r,x=0)=\frac{\Delta w^2\Delta u}{2\sqrt{2\pi}}
\frac{1}{r},\quad r\gg1.
\end{equation}
It is possible to show that for the inviscid case ($\delta_u\to 0,\ \delta_\rho\to0 $), where the velocity and density distributions become discontinuous and can be approximated with the Heaviside $\theta$-function $u\sim\theta(x)$, the local contribution of the dissipation scales as $D_{wwu}\sim r^{-1}$ near the discontinuity. However, the mean anomalous energy dissipation term remains finite,
\begin{equation}
    \langle D_{wwu}\rangle\sim 1 \,.
\end{equation}
In this sense, a Burgers-type shock layer provides a simple one-dimensional model for anomalous dissipation.

\subsubsection{Randomly oriented 3D shock layer model and comparison with DNS}
\begin{figure}
	\begin{center}
		\includegraphics[width=0.5\linewidth]{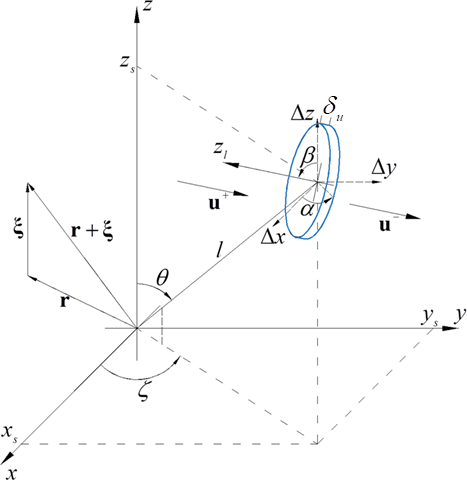}
	\end{center}
	\caption{ Burgers shock-layer in the laboratory coordinate system. The parameters $(x_s, y_s,z_s)$, which can be written in spherical coordinates $(l,\zeta,\theta)$, determine the center point of the shock layer. The angles $(\alpha,\beta,\gamma)$ determine its orientation. Vector $\bm r = (x, y,z)$ represents the reference point and vector $\bm \xi = (0, 0, \xi_z)$ represents the increment vector. The longitudinal velocity increment is defined as $\delta_\xi u_z = [\bm u(\bm r+\bm \xi)-\bm u(\bm r)]\cdot \bm \xi/|\bm \xi|=\delta_{||}u$.}
	\label{fig:randomly_oriented_shock}
\end{figure}

In the following, we increase the complexity of the shock-layer model to further approach the situation in a turbulent Navier-Stokes flow. In the next section, which will discuss the geometry of high-dissipation regions, we will show that their spatial support is found in the form of curved sheets; this suggests a kinematic model for this layer geometry. To this end, the results from the last subsection are generalized exactly along these lines. We follow the ideas of \citet{Kambe_2000} and \citet{Kambe_1997}; see also \citet{Zinchenko_2024}. The authors developed kinematic turbulence models based on randomly oriented Burgers-type structures. We extend the one-dimensional shock-layer profile of the last subsection to a single randomly oriented shock-layer in three-dimensional space as illustrated in figure \ref{fig:randomly_oriented_shock}.
Let ${\bm x}_s$ be the center of the shock-layer and let ${\Delta \bm x}=\bm x-\bm x_s$ denote the shift in the laboratory coordinate system,
\begin{equation}
    x_s=l\cos\zeta\sin\theta,\quad y_s=l\sin\zeta\sin\theta,\quad 
    z_s=l\cos\theta.
\end{equation}
The orientation of the local frame is described by an orthogonal Euler rotation matrix $M_R=R_z(\alpha)R_x(\beta)R_z(\gamma)$, which is composed of three elemental three-dimensional rotations. The local and laboratory coordinates are then related by
\begin{equation}
    \bm x_l=M_R^T\Delta \bm x
\end{equation}
Therefore, a velocity field prescribed in the local coordinate system $\bm u_l(\bm x_l)$ is transformed to the laboratory frame as
\begin{equation}
    \bm u_{\rm lab}(\bm x) =M_R\bm u_l(M_R^T\Delta \bm x).
\end{equation}
For the rotation matrix $M_R$, angle $\gamma$ can be set to zero as the Burgers shock layer is symmetric in the tangential plane. Thus, the simplified rotational matrix is
\begin{equation}
M_R(\alpha,\beta)=
    \begin{pmatrix}
    \cos\alpha & -\sin\alpha\cos\beta & \sin\alpha\sin\beta \\
    \sin\alpha & \cos\alpha\cos\beta & -\cos\alpha\sin\beta \\
    0 & \sin\beta & \cos\beta
    \end{pmatrix}.
\end{equation}
We now compare this analytical shock-layer model with local shock events identified in the DNS data by isolating high-amplitude structures in homogeneous isotropic compressible turbulence and comparing them with the analytical model in the Duchon-Robert framework in physical and scale space.

In the present case, we select one isolated shock-layer event. The center $\bm x_s$ is chosen as the average position of all points belonging to the shock event. The parameters of the analytical profile, $\bar u$, $\Delta u$, $\delta_u$, $\bar\chi$, $\Delta\chi$, and $\delta_\rho$, are obtained from one-dimensional cuts through the event along the local normal direction. In particular, the numerical profiles are compared with \eqref{eq:burgers_shock}.

\begin{figure}
	\begin{center}
		\includegraphics[width=0.9\linewidth]{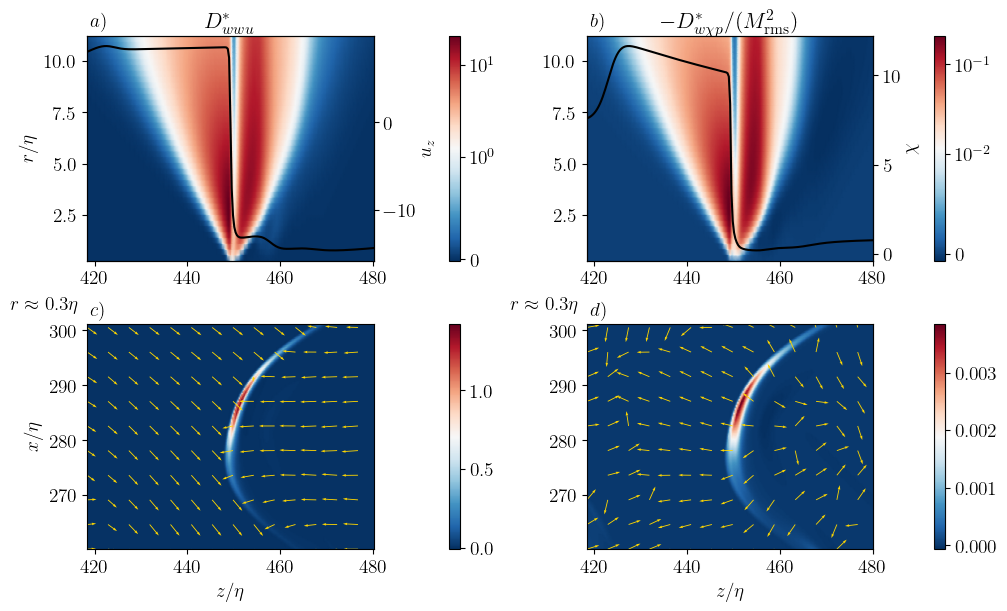}
	\end{center}
	\caption{Dissipation terms obtained from the DNS data. Panels a) and c) show the normalized deformation work, $D_{wwu}^*=D_{wwu}/(\|w\|^2_\infty\,\|u\|_\infty)$, with panel a) in coarse-graining space and c) in physical coordinate space. Panels b) and d) show the normalized baropycnal work, $-D_{w\chi p}^*=-D_{w\chi p}/(\|w\|_\infty\,\|\chi\|_\infty\,\|p\|_\infty)$, with panel b) in coarse-graining space and d) in physical coordinate space. The black lines in a) and b) display the corresponding profiles of $u_z$ and $\chi$, respectively. Yellow vectors in panels c) and d) highlight velocity vector field $\bm u$ in the plane and the gradient of the inverse square-root density $\bm \nabla \chi$, respectively. Data are again for $Re=2400$.}
	\label{fig:diss_num}
\end{figure}

\begin{figure}
	\begin{center}
		\includegraphics[width=0.9\linewidth]{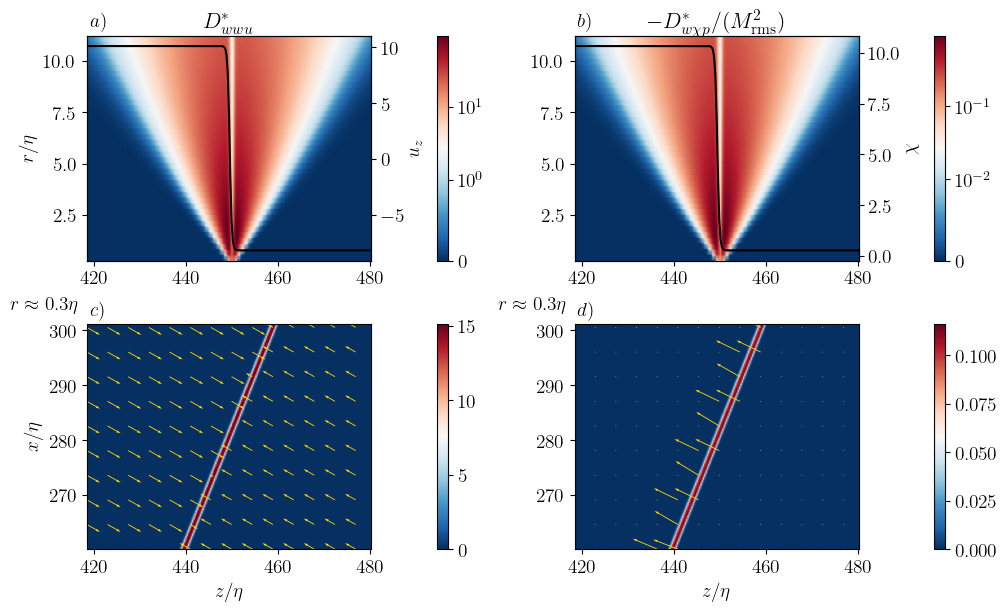}
	\end{center}
	\caption{Dissipation terms obtained from the analytical data for a single shock layer of the analytical model with parameters matched to those of the DNS data in figure \ref{fig:diss_num} for comparison. Panels a) and c) show the normalized deformation work, $D_{wwu}^*$, and panels b) and d) show the normalized baropycnal work, $-D_{w\chi p}^*$.}
	\label{fig:diss_analyt}
\end{figure}

For the local DR analysis, we take a longitudinal increment in the laboratory coordinate system, in particular, we take vertical increments $\bm \xi=(0,0,\xi_z)$. The analytical longitudinal velocity increment is then
\begin{equation}
\label{eq:vel_incr}
    \delta_\xi u_z = \left[\bm u_{\rm lab}(\bm x+\bm \xi) -\bm u_{\rm lab}(\bm x)\right]\cdot \frac{\bm \xi}{|\bm \xi|} =u_z(\bm \Pi, x,y,z+\xi_z)-u_z(\bm \Pi,x,y,z).
\end{equation}
where $\bm \Pi= (\alpha,\beta,\gamma,\theta,\zeta, l, \delta_u,\Delta u, \bar u)$ determines the position and direction of the shock-layer in the laboratory coordinate system, as well as the thickness and amplitudes. In its full form, the velocity increment is defined as
\begin{equation}
    \delta_\xi u_z=\Delta u \cos \beta\left[\tanh\left(\frac{s}{\delta_u}\right)-\tanh\left(\frac{s+\xi_z\cos \beta}{\delta_u}\right)\right],
\end{equation}
with $s=\sin\alpha\sin\beta x-\cos\alpha\sin\beta y+\cos\beta z$. The expressions for $\delta_\xi \chi$, $\delta_\xi w$ and $\delta_\xi p$ are obtained analogously. The analytical and numerical dissipation contributions are then calculated from \eqref{eq:diss_terms} with the test function defined in \eqref{eq:test_function}. In figures \ref{fig:diss_num} and \ref{fig:diss_analyt}. They compare the DR dissipation contributions obtained from the DNS data and from the analytical Burgers shock-layer model. 

The analytical model captures the main features of the high-amplitude dissipation event of the simulation fairly well as seen by comparison of figures \ref{fig:diss_num} and \ref{fig:diss_analyt}. The deformation work term $D_{wwu}$ and the baropycnal work term $D_{w\chi p}$ are localized near the shock layer. The differences between the models arise from the idealized one-dimensional nature of the analytical model. The Burgers-type shock layer would be infinitely extended in the tangential directions. It is straight, isolated, and described by smooth one-dimensional profiles. In contrast, the shock-layer event observed in the DNS is finitely extended and curved. Consequently, dissipation fields from DNS are less symmetric and are more strongly affected by surrounding fluctuations and by the finite size of the structure. Nevertheless, the comparison shows that a single Burgers-type shock layer provides a well-fitted local model for the dominant DR dissipation structures observed in the DNS at the very high Mach numbers.

\begin{figure}
    \begin{center}
        \includegraphics[width=0.95\linewidth]{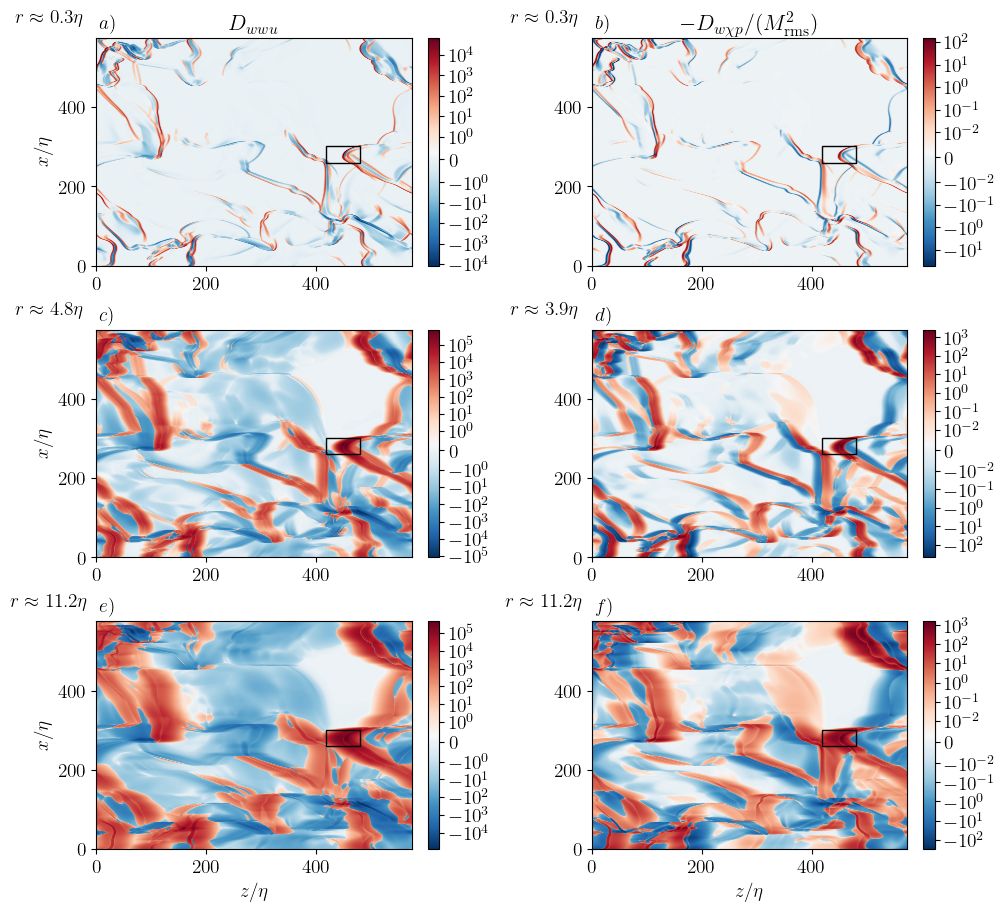}
    \end{center}
    \caption{Dissipation terms in the $x-z$ plane for different coarse-graining scales. Panels a), c), and e) show the deformation work $D_{wwu}$, and panels b), d), and f) show the baropycnal work $-D_{w\chi p}$. The black square marks the local shock event from figure \ref{fig:diss_num}.}
    \label{fig:local_dissiaption}
\end{figure}

To show how the selected shock-layer event looks in the full cross section, we additionally compute the two anomalous dissipation terms in the $x-z$ plane. Figure \ref{fig:local_dissiaption} shows the corresponding fields in contour slice cuts. Deformation $D_{wwu}$ (left) and baropycnal work contributions $-D_{w\chi p}$ (right)  are displayed for three different coarse-graining scales $r$: the first in the viscous subrange, the second between viscous and inertial range (where dissipation reaches its maximum), and the third in the inertial subrange. 

At the smallest scale $r$, the dissipation is concentrated in thin, sharp structures, while at larger $r$ the same structures become broader and more spatially correlated. For this particular snapshot, the volume-averaged deformation work $\langle D_{wwu}(r)\rangle_V$ reaches its maximum at $r_M/\eta\approx 4.8$, and the volume-averaged baropycnal work at $r_M/\eta\approx3.9$. The characteristic scale at which the mean dissipation is maximal seems to be slightly different for the deformation and baropycnal terms. The transition toward inertial-range behavior occurs beyond these scales $r>r_M$, and for the baropycnal work, it seems to start at smaller scales.

\begin{figure}
	\begin{center}
		\includegraphics[width=0.7\linewidth]{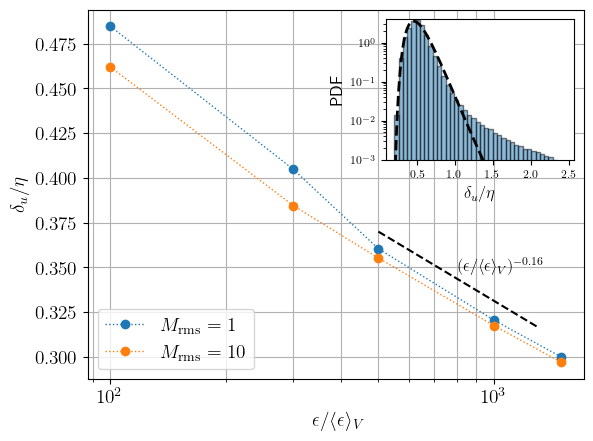}
	\end{center}
	\caption{The most probable shock layer thicknesses for $M_{\rm rms}=1$ and $M_{\rm rms}=10$ as functions of the normalized dissipation threshold $\epsilon/\langle\epsilon\rangle_V$. The inset shows the probability density function (PDF) for $M_{\rm rms}=1$ at $\epsilon/\langle \epsilon\rangle_V=10^2$. The average thickness is calculated over all events in the computational domain. The dashed line shows a lognormal distribution approximation of the mean PDF.}
	\label{fig:shock_thickness}
\end{figure}

Next, the anomalous dissipation terms are determined for a randomly oriented Burgers shock layer at a certain distance. The homogeneous isotropic statistics of the turbulence suggest an averaging over all directions of the shock layer for this case, i.e., over all possible combinations of angles $\alpha$, $\beta$ and $\gamma$. We thus obtain
\begin{equation}
    \langle D_{wwu}(r,\bm x)\rangle_{\alpha,\beta}=\frac{1}{4r^2}\frac{1}{4\pi}\int_{-\infty}^{+\infty}\int_{0}^{\pi}\int_{0}^{2\pi}\psi_1\left(\frac{\xi_z}{r}\right)(\delta_\xi w_z)^2\delta_\xi u_z\,\sin\beta\, d\alpha\, d\beta\, d\xi_z.
\end{equation}
If the center of the shock-layer is randomly located within a sphere of radius $l$ around the origin of the laboratory coordinate system, the dissipation term must additionally be averaged over the angles $\theta$ and $\zeta$. Moreover, following \citet{Kambe_2000}, the probability distribution of the distance $l$ is taken into account, as illustrated again in figure \ref{fig:randomly_oriented_shock}. It is described by the probability density function
\begin{equation}
P_l(V_l)=\frac{1}{\langle V\rangle}\exp\left( -\frac{V_l}{\langle V\rangle}\right).
\end{equation}
As a simple model, only the nearest shock layer is assumed to dominate the velocity difference. The nearest point of the shock to the origin is therefore assumed to be distributed according to a Poisson statistical ensemble, which describes discrete statistical events. With the sphere volume $V_l=4\pi l^3/3$, the following distribution follows for $l$:
\begin{equation}
    P_l(l)=3bl^2\exp(-bl^3)\quad \text{with} \quad b=\frac{\Gamma^3(4/3)}{l_0^3}.
\end{equation}
Second, we have to consider an ensemble of shocks of various thicknesses. According to figure \ref{fig:shock_thickness}, we take a lognormal distribution of shock thicknesses,
\begin{equation}
    P_\delta(\delta_u)=\frac{1}{\delta_u\sigma_\delta\sqrt{2\pi}}
\exp\left[-\frac{\ln^2\delta_u/\delta_0}{2\sigma_\delta^2}\right],
\end{equation}
where $\delta_0$ is the median, or characteristic, value of the shock-layer thickness. Thus, half of the probability lies below $\delta_0$ and half above it. The parameter $\sigma_\delta$ is the standard deviation of $\ln(\delta_u/\delta_0)$. Physically, $\sigma_\delta$ controls the spread of the shock-layer thickness, larger values of $\sigma_\delta$ correspond to a broader distribution. 

Figure \ref{fig:shock_thickness} shows that the most probable shock-layer thickness decreases systematically as the local dissipation threshold increases. For both rms Mach numbers, the characteristic thickness is below the mean Kolmogorov scale $\delta_u/\eta<1$ for the threshold $r/\langle r \rangle_V=10^2$. Note that these scales are resolved in the DNS as we use $k_{\rm max}\eta > 10$. The nondimensionalized thickness $\delta_u/\eta$ weakly depends on the Mach number. For the $M_{\rm rms}=1$ case, we observe slightly larger most-probable thicknesses than for $M_{\rm rms}=10$, but both curves follow a similar power-law scaling. The inset illustrates the full probability density of $\delta_u/\eta$ for $M_{\rm rms}=1$ at $r/\langle r\rangle_V=10^2$. This distribution can be approximated by the lognormal distribution shown by the dashed curve.

\begin{figure}
	\begin{center}
		\includegraphics[width=0.8\linewidth]{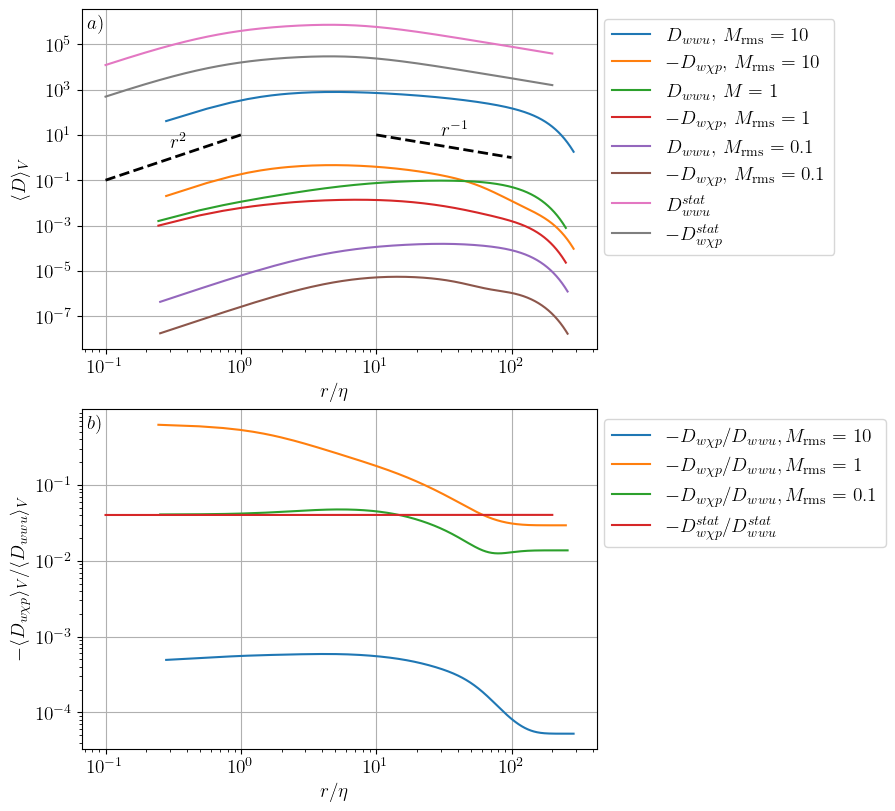}
	\end{center}
	\caption{Volume-averaged anomalous dissipation contributions arising in the DR framework as functions of the normalized coarse-graining scale $r/\eta$. DNS results are shown for  $M_{\rm rms}=0.1, 1$ and 10. Here, $\langle D^{\rm DNS}_{wwu}\rangle_V$ is the deformation-work and $-\langle D^{\rm DNS}_{w\chi p}\rangle_V$ is the baropycnal-work. Curves of $\langle D^{\rm stat}_{wwu}\rangle_{\bm \Pi,V}$ and $-\langle D^{\rm stat}_{w\chi p}\rangle_{\bm\Pi,V}$ show the corresponding statistical predictions for the randomly oriented Burgers-type shock-layer ensemble. Panel a) shows the actual mean dissipation. Dashed black lines indicate power-law scalings. Panel b) shows the ratio $\langle D_{w\chi p}\rangle_V/\langle D_{wwu}\rangle_V$ for all Mach numbers and statistical case.}
	\label{fig:mean_diss}
\end{figure}

With the given distributions, the dissipation term results in
\begin{equation}
    \langle D_{wwu}(r,\bm x)\rangle_{\bm \Pi}=\frac{1}{4\pi}\int_0^\infty\int_0^\infty\int_0^{2\pi}\int_0^\pi\langle D_{wwu}(r,\bm x)\rangle_{\alpha,\beta}P_l(l)P_\delta(\delta_u)\sin\theta\, d\theta\, d\zeta\, dl\, d\delta_u,
    \label{eq:diss_scaling}
\end{equation}
where $\bm \Pi=(\alpha,\beta,\theta,\zeta,l,\delta_u)$ denotes the set of random shock-layer parameters, including the orientation, position, and thickness distributions. The averaged dissipation term can be also expressed in terms of a mixed third-order structure function as
\begin{equation}
\langle D_{wwu}(r)\rangle_{\bm \Pi,V}
=\frac{1}{4r^2}\int_{-\infty}^{+\infty}
\psi_1\left(\frac{\xi}{r}\right)
\langle S_{wwu}(\xi)\rangle_{\bm \Pi,V}\,
d\xi .
\label{eq:mean_diss_term}
\end{equation}
with
\begin{equation}
    \langle S_{wwu}(\xi)\rangle_{\bm \Pi,V}=\frac{1}{16\pi^2V}\int_{\bm \Pi}\int_{V} \delta_\xi w_z^2 \delta_\xi u_z\, d\bm \Pi \, dV.
\label{eq:str_func}
\end{equation}
In figure \ref{fig:mean_diss}, we compare dissipation terms $D_{wwu}$ and $D_{w\chi p}$ for statistical kinematic model \eqref{eq:mean_diss_term}, \eqref{eq:str_func} and for DNS data with $M_{\rm rms}=(0.1,1,10)$. Figure \ref{fig:mean_diss} shows the corresponding mean DR dissipation terms \eqref{eq:diss_deform} and \eqref{eq:diss_barop} and their ratio for DNS data and for the kinematic model following \eqref{eq:mean_diss_term}. Generally, these terms are obtained by averaging local dissipation over full volume; see, for example, figure~\ref{fig:local_dissiaption} for the local distribution of the dissipation terms for different coarse-graining scales. At small coarse-graining scales $r/\eta<1$, the magnitude of the dissipation vanishes approximately with $r^2$. This agrees with the analytical expectation for smooth fields in the viscous subrange. At intermediate scales $r$, the curves reach a maximum at a few Kolmogorov lengths. At larger scales $r$, the dissipation decays approximately as $r^{-1}$, in agreement with the Burgers shock-layer prediction. 

For $M_{\rm rms}=10$, dissipation has the largest amplitudes, followed by $M_{\rm rms}=1$, while the $M_{\rm rms}=0.1$ contributions are much smaller. In the DNS data, the deformation term contribution $D_{wwu}$ is larger than the baropycnal one, $D_{w\chi p}$. This holds for all Mach numbers. The statistical model captures a scaling very similar to that of the compressible homogeneous isotropic DNS. It shows that the model of a random ensemble of Burgers-type shock layers provides a useful description of the anomalous-dissipation structures. Panel b) of figure \ref{fig:mean_diss} shows that the biggest influence of the baropycnal work lies in the range of moderate Mach numbers $M_{\rm rms}\sim1$. For all rms Mach numbers, deformation work is dominant at all scale, i.e., $\langle D_{wwu}\rangle_V\gg -\langle D_{w\chi p}\rangle_V$. The relatively large contribution of the baropycnal term at $M_{\rm rms}=1$ may help explain why the mean dissipation in figure \ref{fig:dissi_anomaly} remains on the lower branch, even though shocks have already emerged. Since $\langle D_{w\chi p}\rangle_V<0$, the baropycnal term partially cancels the deformation contribution, thereby reducing the total mean anomalous dissipation. For $M_{\rm rms}=10$, in contrast, the deformation work is dominant and the relative baropycnal part is negligible, so it does not compensate the deformation work. This may explain why the data are found in the second branch.

\section{Geometric analysis of energy dissipation fields}
\label{sec:multi}

\begin{figure}
    \centering
    \includegraphics[width=0.8\linewidth]{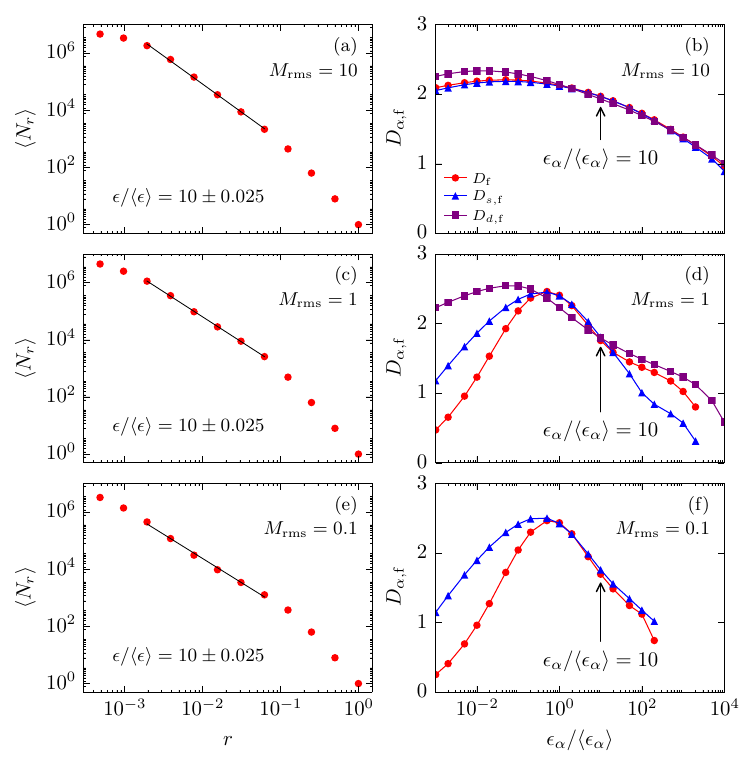}
    \caption{Box-counting dimensions of the dissipation fields at rms Mach numbers $M_{\rm rms}=10$, $1$, and $0.1$ are shown in the top, middle, and bottom rows, respectively. The mean number of cubic boxes, $\langle N_r\rangle$, versus box size $r$ for the total dissipation field at an isolevel $\epsilon / \langle \epsilon \rangle=10 \pm 0.025$ is plotted in a double-logarithmic plot in the left column panels (a), (c), and (e). The black solid lines represent the power-law fits, whose slopes provide the corresponding box-counting dimensions $D_{\rm f}$. Panels (b), (d), and (f) in the right column show the box-counting dimensions $D_{\alpha,\mathrm f}$, where $\alpha=\{\cdot,s,d\}$, for the total, solenoidal, and dilatational dissipation fields at different isolevels $X=\epsilon_\alpha / \langle\epsilon_\alpha\rangle$, using a $\pm0.025$ margin around each isolevel. The Reynolds number $Re=2400$.}
    \label{fig:Nr_Df}
\end{figure}
Randomly positioned and oriented Burgers-type shock layers provide a useful model of anomalous dissipation in the highly supersonic regime. The anomalous-dissipation fields predicted by the model agree qualitatively with those of an isolated DNS shock layer, although the analytical layer is planar and infinitely extended, whereas the DNS layer is curved and has a finite extent. The volume-averaged contributions also exhibit similar scale dependence in the model and the DNS. This shock-layer organization is consistent with that of the nonlinear production fields of the dissipation rates, which become concentrated within dominant shock layers in the highly supersonic regime. As these production fields change their spatial organization with Mach number, so does the geometry of the dissipation-rate fields. Figure~\ref{fig:isosurface} qualitatively illustrates this correspondence. At $M_{\rm rms}=0.1$, regions of intense total and solenoidal dissipation form patchy shear layers and vortex tubes, respectively, whereas at $M_{\rm rms}=10$, regions of intense total and dilatational dissipation nearly coincide in curved, approximately two-dimensional sheets. We use box-counting and multifractal analyses to quantify this Mach-number-dependent geometric change in the following.

\subsection{Box-counting dimension of isolevel sets}
We compute the box-counting dimensions $D_{\alpha,{\rm f}}$ of the total, solenoidal and dilatational dissipation fields $\epsilon_\alpha$ with $\alpha=\{\cdot,s,d\}$ by the algebraic power law
\begin{equation}
\langle N_r\rangle \sim r^{-D_{\alpha,\rm f}} \,,
\end{equation}
with the mean number of cubic boxes of side length $r$ that completely cover an isolevel set of the field of interest \citep{SS2006}. See the left column of figure \ref{fig:Nr_Df} for $M_{\rm rms}=0.1, 1$, and 10 at $Re=2400$. The mean is taken over different snapshots of the flow. The fractal dimensions are determined for the range of $10^{-3} \leq \epsilon_\alpha / \langle \epsilon_\alpha \rangle \leq 10^{4}$ to characterize the geometry of regions with weak to intense dissipation; see the right column of figure~\ref{fig:Nr_Df}. The slopes were evaluated for the same range of scales as indicated in the panels in the left column. A margin of $X\pm0.025$ around each isolevel $X$ is used for the box-counting dimension determination. 

For all three rms Mach numbers, the box-counting dimension is close to 2 for isolevel sets at the corresponding mean dissipation. In the case of $M_{\rm rms}=10$, a wide range of isolevels appears in the form of sheets with $D_{\alpha,\rm f}\approx 2$, except the largest amplitudes. The box-counting dimensions for all three fields $\epsilon_{\alpha}$ collapse almost perfectly, which supports the visualizations from figure~\ref{fig:isosurface}.  

\subsection{Multifractal analysis}
We now employ the multifractal formalism to quantify the Mach-number-dependent changes in the effective dimensions of the dissipative structures associated with the two branches of the dissipative anomaly. These changes also manifest themselves as increased intermittency of the dissipation fields \citep{AFS2026}. Intermittency is commonly quantified through the anomalous scaling of velocity structure functions or the scaling of dissipation-rate moments, both of which describe departures from self-similar scaling and provide statistical measures of intermittency. A geometrical characterization of this intermittency is provided by the generalized dimensions of the multifractal formalism. In incompressible turbulence, the multifractal nature of the dissipation field has been extensively investigated and provides the basis for phenomenological descriptions of intermittency \citep{Meneveau:NPB1987,Sreenivasan:ARFM1991,SSY2005,Mukherjee:PRL2024}. Multifractal studies of the dissipation field in compressible turbulence remain scarce. A recent study examined its multifractal properties in fully compressible convection \citep{Alam_2025}.

\begin{figure}
	\begin{center}
	\includegraphics{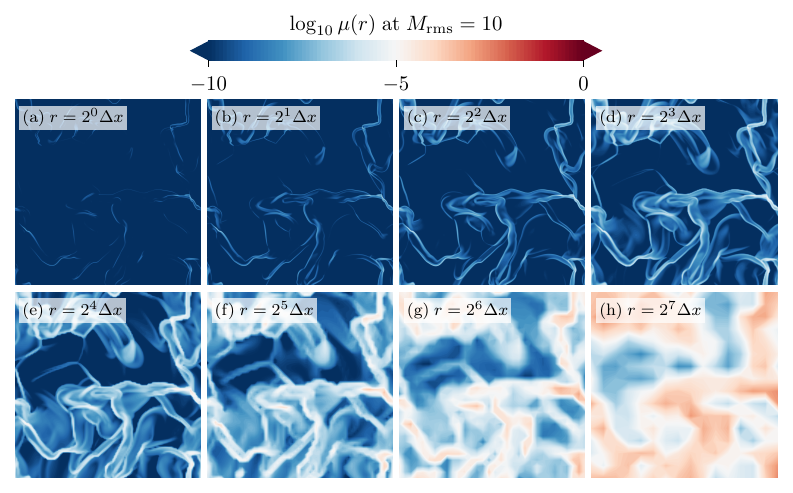}
	\end{center}
	\caption{For $M_{\rm rms}=10$, snapshot contour slices of the coarse-grained measure $\mu(r)$ at coarse-graining scales $r$ from the grid spacing, $\Delta x = L_{\rm box}/2048$, to $2^7 \Delta x$. The data are presented in terms of the decadic logarithm. The Reynolds number $Re=2400$.}
	\label{fig:mu_M_10}
\end{figure}

\begin{figure}
	\begin{center}
		\includegraphics{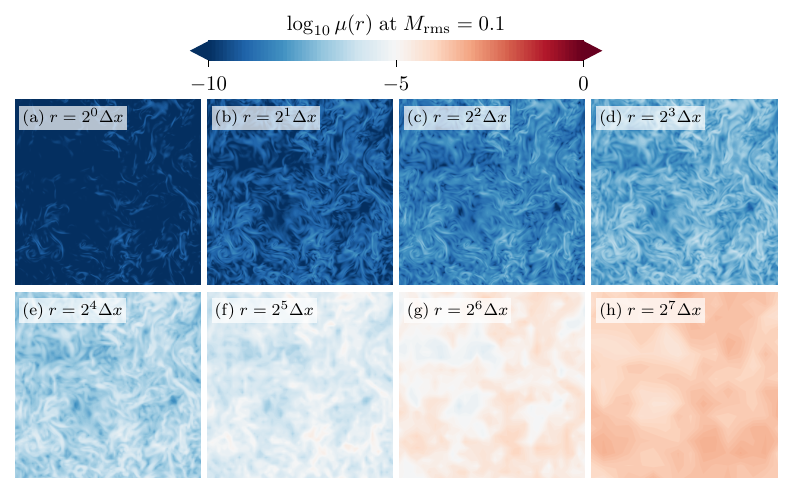}
	\end{center}
	\caption{Same as figure \ref{fig:mu_M_10} for $M_{\rm rms}=0.1$}
	\label{fig:mu_M_0p1}
\end{figure}

To this end, the computational domain is divided into non-overlapping cubic boxes of side length $|{\bm r}| = r$. The coarse-grained dissipation contained within the $i$th cubic box (centered at ${\bm x}_i$) is obtained by
\begin{equation}
{\cal E}_{i}(r) = \int_{B_i(r)} \epsilon({\bm x}_i + {\bm x}')d^3{\bm x}' \quad {\rm with} \quad \bigcup_{i=1}^{N(r)} B_i(r) = V,
\end{equation}
where $B_i(r)$ denotes the $i$th box. The corresponding (normalized) coarse-grained measure is
\begin{equation}
\mu_{i}(r) = \frac{{\cal E}_{i} (r)}{\cal E} \quad \mbox{with} \quad \sum_i\mu_{i}(r)=1.
\end{equation}
Here, ${\cal E}=\langle\epsilon\rangle_VV$ is the total dissipation.
Figures~\ref{fig:mu_M_10} and~\ref{fig:mu_M_0p1} present how the coarse-grained dissipation measure $\mu(r)$ changes with the box size   $r$ in the highly compressible case, $M_{\rm rms}=10$, and nearly incompressible case, $M_{\rm rms}=0.1$, respectively.
At $M_{\rm rms}=10$, figure~\ref{fig:mu_M_10} shows a strong spatial separation between regions of intense and weak dissipation. At small box sizes, this strong spatial contrast is largely preserved: boxes intersecting the shock layers contain large values of $\mu(r)$, whereas neighbouring boxes remain several orders of magnitude smaller. As the box size increases, the shock layers broaden because each box averages over a larger neighbourhood. Nevertheless, the measure remains concentrated along the same shock-associated structures over a wide range of scales. Only at the largest box sizes, individual boxes encompass both the shock-associated layers and the surrounding weakly dissipative regions, leading to a rapid homogenization of the measure as $r$ approaches the domain size.

For $M_{\rm rms}=0.1$, figure~\ref{fig:mu_M_0p1} shows a markedly different spatial organization. The dissipation is distributed among numerous small-scale shear layer structures throughout the domain, with weakly and strongly dissipative regions closely interspersed. Consequently, increasing the box size modifies the local measure more gradually. The fine-scale structures progressively merge with their surroundings, and the spatial variations decrease continuously with increasing $r$, without the pronounced large-scale transition observed for $M_{\rm rms}=10$. At the largest scales, both cases approach a uniform measure, consistent with the normalization $\sum_i\mu_i(r)=1$.

The distinct spatial organizations of the coarse-grained measure discussed above are quantified by the scaling of its moments,
\begin{equation}
\sum_i\mu_i^q(r)\sim r^{(q-1)D(q)}\,.
\end{equation}
The moments are then re-expressed as the partition function 
\begin{equation}
\Xi(q,r)=\left[\sum_i\mu_i^q(r)\right]^{\frac{1}{q-1}}
\end{equation}
to obtain $D(q)$, the spectrum of generalized dimensions, which are obtained as~\citep{Hentschel:PD1983}
\begin{equation}
D(q) = \lim_{r\rightarrow0} \frac{1}{q-1} \dfrac{\log \sum_i\mu_i^q(r)}{\log r}
\label{eq:Dq}
\end{equation}
Different moment orders $q$ emphasize different parts of the dissipation measure. Positive values of $q$ preferentially weight boxes containing the largest values of $\mu(r)$ and therefore probe the shock- or shear-associated intense dissipative structures, while negative values emphasize boxes with small $\mu(r)$ and thus probe the weakly dissipative regions. The generalized dimensions $D(q)$ quantify the effective fractal dimensions of the structures that dominate the measure at each order.
$D(q=0)$ is in principle a box-counting dimension, here however always $D(0)=3$ since the whole volume is considered and not isolevel subsets.
\begin{figure}
	\begin{center}
		\includegraphics[width=0.85\linewidth]{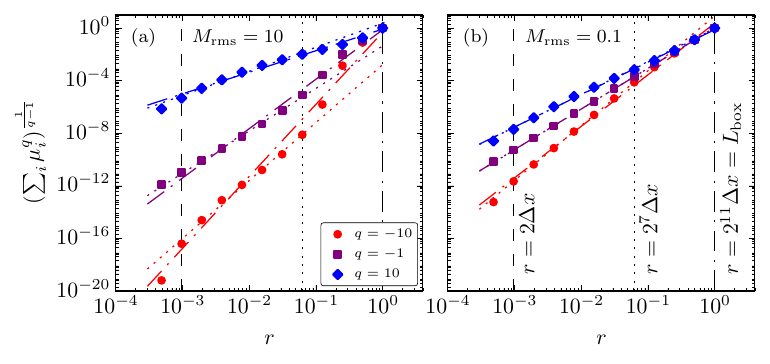}
	\end{center}
	\caption{Double logarithmic plots of the partition function $\Xi(q,r)$ versus $r$. Data are shown for powers of $q = \{-10, -1, 10\}$ to illustrate the scaling range for the evaluation of the corresponding generalized dimensions $D(q)$ by a power-law fit. The vertical black lines mark $r=2\Delta x$, $r=2^7\Delta x$, and $r=L_{\rm box}$. The dotted and dash-dotted lines represent power-law fits over $2\Delta x \leq r \leq 2^7\Delta x$ and $2\Delta x \leq r \leq L_{\rm box}$, respectively. (a) $M_{\rm rms}=10$ and (b) $M_{\rm rms}=0.1$.  The Reynolds number $Re=2400$.}
	\label{fig:partition_function}
\end{figure}

Figure~\ref{fig:partition_function} shows $\Xi(q,r)$ as a function of the coarse-graining scale $r$ for orders $q={-10,-1,10}$ at $M_{\rm rms}=10$ in panel (a) and $M_{\rm rms}=0.1$ in panel (b). The slopes of the double-logarithmic curves provide the corresponding generalized dimensions through \eqref{eq:Dq}. For $M_{\rm rms}=0.1$, the curves vary comparatively smoothly with $r$. This behavior reflects the close spatial interspersion of strongly and weakly dissipative regions shown in figure~\ref{fig:mu_M_0p1}. For $M_{\rm rms}=10$, a more pronounced change of slope appears at the larger coarse-graining scales. This behavior follows from the strong separation between the intense dissipation concentrated along the dominant shock layers and the broad weakly dissipative regions between them. As box size (or coarse-graining scale) $r$ becomes sufficiently large, coarse-graining progressively averages the dissipation within the shock layers together with that in the surrounding weakly dissipative regions. The contrast between neighbouring boxes therefore decreases. As a result, the large positive-order moments flatten. Conversely, boxes that initially contain little dissipation progressively include portions of the neighbouring shock layers. Their increasing measure gives rise to the upturn of the large negative-order moments visible in figure~\ref{fig:partition_function}(a).

\begin{figure}
	\begin{center}
		\includegraphics[width=0.8\linewidth]{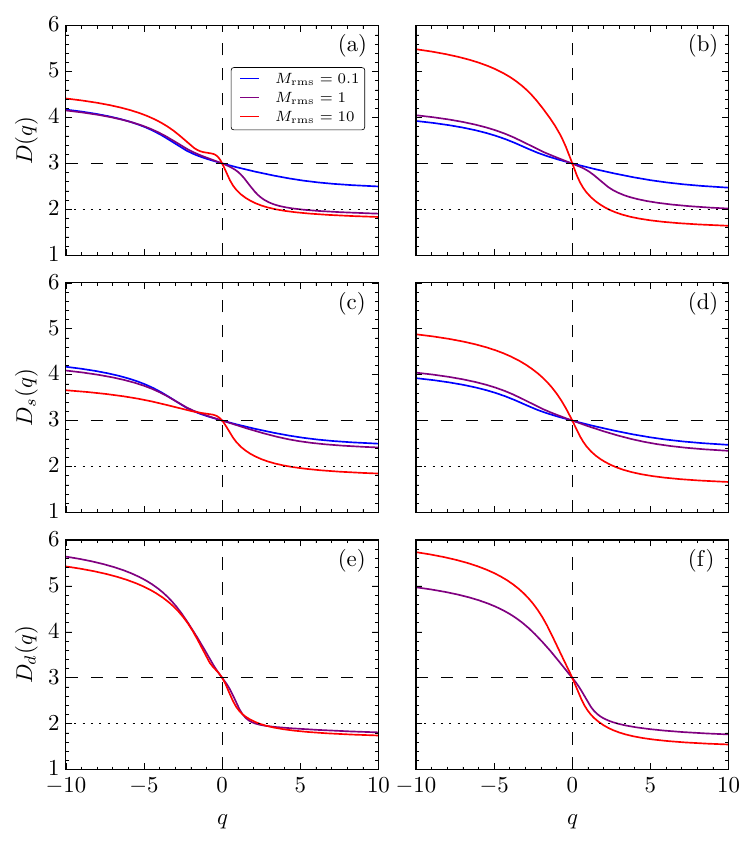}
	\end{center}
	\caption{Generalized dimensions of the total dissipation rate field and its components for $M_{\rm rms} = 0.1, \ 1$ and $10$ at $Re \approx 2400$. Panels (a, b), (c, d), and (e, f) show $D(q)$, $D_s(q)$, and $D_d(q)$, respectively. The dimensions are obtained from power-law fits over the ranges $r \in [2\Delta x,\,2^7\Delta x]$ in the left column and $r \in [2\Delta x,\,L_{\rm box}]$ in the right column.}
	\label{fig:Dq}
\end{figure}
The generalized dimensions of the total $D(q)$, solenoidal $D_s(q)$ and dilatational  $D_d(q)$ dissipation fields are shown in figure \ref{fig:Dq} for the DNS runs at $Re=2400$. We determine them using two fitting intervals. In the left column, we fit the moments over the range $2\Delta x \le r \le 2^7\Delta x$, which emphasizes the small- and intermediate-scale behavior. In the right column, we extend the fitting range to $2 {\rm \Delta} x \le r \le L_{\rm box}$, which additionally includes the large-scale crossover. Comparison of the two therefore provides a measure of the sensitivity of the inferred dimensions to the large-scale spatial organization of the dissipation field. The main observations can be summarized as the following points:

(1) The generalized dimensions of the total dissipation field, $D(q)$, are shown in panels (a) and (b) of figure~\ref{fig:Dq}. For $M_{\rm rms}=0.1$ and $1$, $D(q)$ changes only modestly when the fitting interval is extended to $r=L_{\rm box}$. In contrast, the $M_{\rm rms}=10$ case is much more sensitive to the fitting range, particularly for large $|q|$. This sensitivity results from the pronounced large-scale crossover discussed above.

(2) For positive orders, which preferentially sample the intense dissipation, the generalized dimensions approach values close to $D(q)\simeq2$ for $M_{\rm rms}=1$ and $10$ in panel (a). A value of $D(q) \approx 2$ indicates that the strongest dissipation is concentrated on a subset of the flow with an effective dimension close to two. This is consistent with the organization of the intense dissipation around the shock layers observed in the visualizations. The spatial organization nevertheless differs between the two cases. At $M_{\rm rms}=1$, the intense dissipation is associated primarily with fragmented shocklets. At $M_{\rm rms}=10$, it is concentrated around a small number of dominant shock layers. For $M_{\rm rms}=0.1$, $D(q)$ approaches a larger value of approximately $2.5$, indicating that the strongest dissipation occupies a more space-filling network of small-scale shear-layer structures. The similarity of the negative-order dimensions for $M_{\rm rms}=1$ and $0.1$ indicates that the presence of fragmented shocklets at $M_{\rm rms}=1$ does not alter the close interspersion of weakly and strongly dissipative regions characteristic of the flow.

(3) The generalized dimensions of the solenoidal dissipation field, $D_s(q)$, in panels (c) and (d) are nearly identical for $M_{\rm rms}=0.1$ and $1$ over the entire range of $q$. They are also only marginally affected by extending the fitting range. For positive orders, $D_s(q)$ approaches approximately $2.5$ in both cases. The similarity of the generalized dimensions indicates that the emergence of shocklets at $M_{\rm rms}=1$ has little influence on the spatial organization of the solenoidal dissipation, which remains close to that of the nearly incompressible flow. This is consistent with the analysis in section~4, where solenoidal vortex stretching remains dominant and the familiar incompressible-like vortical structures persist up to $M_{\rm rms}=1$. At $M_{\rm rms}=10$, $D_s(q)$ exhibits trends similar to those of the total dissipation field. For positive orders in panel (c), $D_s(q)$ approaches $2$. This behavior is consistent with the localization of the strongest solenoidal dissipation around the dominant shock layers.

(4) The generalized dimensions of the dilatational dissipation field, $D_d(q)$, in panels (e) and (f) show a distinct dependence on the fitting range at $M_{\rm rms}=1$. For positive orders, $D_d(q)$ remains close to $2$, indicating that the strongest dilatational dissipation is concentrated on a subset with an effective dimension close to two. This is consistent with the fragmented shocklets observed at this Mach number. For negative orders, however, $D_d(q)$ decreases when the fitting range is extended. The weakly dissipative regions are closely interspersed with the fragmented shocklets. Consequently, boxes of small and intermediate size already encompass nearby shocklet contributions, whereas further enlargement adds proportionally less dilatational measure. The resulting reduction in the growth rate of the negative-order moments at large $r$ leads to the smaller values of $D_d(q)$ in panel (f). At $M_{\rm rms}=10$, $D_d(q)$ exhibits behavior similar to that of the total and solenoidal dissipation fields. The positive-order dimensions approach $D_d(q) \approx 2$, while the sensitivity to the extended fitting range reflects the pronounced large-scale crossover discussed above.

(5) At $M_{\rm rms}=1$, the weak regions of the total dissipation field remain interwoven with the distributed shear structures, and the isolated shocklets do not substantially modify the weak regions that dominate the negative-order dimensions. By contrast, the weak regions of the dilatational dissipation field remain closely interspersed with the isolated shocklets, giving rise to the stronger fitting-range dependence of the negative-order generalized dimensions discussed above.

Finally, the overall Mach-number dependence of the generalized dimensions complements the intermittency behavior obtained from moments of the energy dissipation rate in  \citet{AFS2026}. The nearly overlapping $D_s(q)$ curves at $M_{\rm rms}=0.1$ and $1$ mirror the weak Mach-number dependence of the solenoidal moment scaling up to $M_{\rm rms}\simeq 1$. The positive-order dimensions of the total and dilatational dissipation at $M_{\rm rms}=1$ are already close to those at $M_{\rm rms}=10$, indicating a similar degree of localization of the intense dissipative events. A similar transition is found in the dissipation-rate moments, with the total and dilatational dissipation approaching their high-Mach-number scaling near $M_{\rm rms}\simeq 1$.
    
\section{Summary and outlook}
\label{sec:con}
In this work, we investigate the dissipative anomaly of the kinetic-energy dissipation field and its solenoidal and dilatational components using DNS of solenoidally forced, isothermal, homogeneous isotropic turbulence over the Mach-number range $0.1 \leq M_{\rm rms} \leq 10$, spanning regimes from nearly incompressible to highly compressible and supersonic turbulence. The Reynolds numbers of the DNS can reach moderate values of $Re\lesssim 2400$.

Our results indicate an existence of dissipative anomalies in isothermal compressible turbulence on the basis of highly resolved DNS. Specifically, all three rescaled dissipation fields exhibit two distinct branches in their Reynolds-number dependence: one for $M_{\rm rms}\leq 1$ and another for $M_{\rm rms}\gg 1$. This Mach-number-dependent distinction is particularly clear for the total and solenoidal dissipation fields. For the dilatational dissipation field, the low-Mach-number branch is observed in our simulations at $M_{\rm rms}=0.77$ and 1, whereas at lower Mach numbers the dilatational contribution remains negligible due to the nearly incompressible character of the flow under solenoidal forcing. We then analyzed the data from three different perspectives---(i) dissipation generation mechanisms, (ii) compressible flow extension of the Duchon-Robert framework of anomalous dissipation, (iii) geometric analysis of high-dissipation regions---to better understand the observed behavior. 

For each field, the two branches seem to approach distinct constant asymptotic values with increasing Reynolds number, demonstrating different dissipative anomalies in the two Mach-number regimes. These results also indicate the presence of a dissipative anomaly in the dilatational dissipation field, a question that remained open in \citet{JDSJFM2021}. We emphasize, however, that, in contrast to the present study, their DNS included the full energy equation and different ways to drive the box turbulence. The latter circumstance differs in the present case as stated already above. The impact of this more comprehensive physical model remains to be understood and cannot be assessed within the present work. To elucidate the origin of the distinct behavior in the two Mach regimes, we analyze the nonlinear production mechanisms of each dissipation field by separating the contributions from solenoidal and dilatational motions and from their mutual interactions.

The moderate Reynolds numbers accessible in our simulations are the price paid for resolving the dissipation fields deep into the viscous subrange. At present, this limitation prevents us from determining conclusively whether the observed dissipative anomaly is weak or strong. In addition, we find that the collapse of the two dissipation components across different Mach numbers depends sensitively on the choice of normalization.

For $M_{\rm rms}\leq 1$, the total and solenoidal dissipation retain the classical incompressible behavior. In this regime, the production of solenoidal dissipation is governed almost entirely by solenoidal vortex stretching, and the associated vortical structures remain close to those found in incompressible turbulence. At $M_{\rm rms}\sim 1$, however, the dilatational contribution to the production of total dissipation becomes comparable to, or even exceeds, that arising from vortical motions. Nevertheless, the total dissipation remains on the low-Mach-number branch, indicating that the magnitude of the dilatational contribution alone cannot explain the separation between the two branches. The dilatational contribution is associated with fragmented shocklets, which do not substantially alter the predominantly vortical organization of the flow. For the dilatational dissipation itself, the production contains no purely solenoidal contribution and consists only of purely dilatational and mixed terms. At $M_{\rm rms}=0.77$ and 1, the former dominate, whereas the mixed contributions remain weak. The corresponding dissipative anomaly can therefore be associated primarily with shocklet structures.

As the Mach number increases further into the supersonic regime, increasingly strong dilatational effects reorganize the flow from predominantly vortical structures interspersed with fragmented shocklets into a shock-layer-dominated state. In the highly supersonic regime, a small number of dominant shock layers become the primary sites of production for all three dissipation fields. There, contributions from purely dilatational motions and their interactions with solenoidal motions dominate, while purely solenoidal contributions become comparatively weak. This transition to a shock-layer-dominated flow organization underlies the distinct high-Mach-number branch of the dissipative anomaly observed in all three dissipation fields.

A complementary analysis based on the Duchon–Robert framework for anomalous dissipation, recently extended to compressible flows by \citet{zinchenko2026}, identifies the strongest precursors of anomalous dissipation within the shock layers. A kinematic model based on randomly oriented Burgers-type shock layers reproduces the DNS data reasonably well. Within the extended framework, two contributions to anomalous dissipation, the deformation term \(D_{wwu}\) and the baropycnal term \(D_{w\chi p}\), emerge as the dominant terms. Their relative magnitude varies with Mach number, providing further evidence for the existence of two distinct branches of the dissipative anomaly.

The distinct spatial organization of the dissipation fields in the low- and high-Mach-number regimes is further quantified through a comprehensive geometrical analysis. This difference is reflected in the scale dependence of the moments of the coarse-grained dissipation measure shown in figure~\ref{fig:partition_function}. The generalized dimensions from a multifractal analysis, which are inferred from these scalings, provide a quantitative characterization of the underlying dissipative structures. At large positive orders $q$, the generalized dimensions of the total and solenoidal dissipation fields, $D(q)$ and $D_s(q)$, approach approximately 2.5 at $M_{\rm rms}=0.1$, consistent with the relatively space-filling intense dissipative structures characteristic of incompressible-like turbulence. At \(M_{\rm rms}=1\), the solenoidal field retains an effective dimension close to that of the nearly incompressible case, whereas the total and dilatational fields approach a dimension of approximately 2, consistent with the emergence of fragmented shocklets. In the highly supersonic regime, the generalized dimensions of all three fields approach an effective dimension close to 2 at large positive orders $q$, consistent with the concentration of intense dissipation in the dominant shock layers.

The present results motivate further investigations of the dissipative anomaly over a broader range of flow conditions. Future work will consider the inclusion of dilatational forcing to assess its influence on the flow dynamics and the associated structural organization. Thermodynamic effects can also be explored by moving beyond the isothermal framework, for example by adopting a polytropic equation of state with $\gamma>1$ or by solving the full energy equation. Such extensions will help establish how robust the observed branches and their asymptotic behavior are with respect to the physical conditions of compressible turbulence, particularly at higher Reynolds numbers.

\vspace{0.5cm}
\noindent
\textbf{Acknowledgements}

\noindent
S.A.~and J.S.~are supported by the European Union (ERC, MesoComp, 101052786). Views and opinions expressed however are those of the authors only and do not necessarily reflect those of the European Union or the European Research Council. G.Z. is supported by the Priority Programme DFG-SPP 2410 CoScaRa of the Deutsche Forschungsgemeinschaft (DFG). C.F.~acknowledges funding provided by the Australian Research Council (Discovery Projects DP230102280 and DP250101526), and the Australia-Germany Joint Research Cooperation Scheme (UA-DAAD). The authors gratefully acknowledge the computing time made available to them on the high-performance computer Otus [under project hpc-prf-kedrict] at the NHR Center Paderborn Center for Parallel Computing (PC$^2$). This center is jointly supported by the Federal Ministry of Research, Technology and Space and the state governments participating in the National High-Performance Computing (NHR) joint funding program. We further acknowledge high-performance computing resources provided by the Leibniz Rechenzentrum and the Gauss Centre for Supercomputing (grants~pr32lo, pr48pi and GCS Large-scale project~10391), the Australian National Computational Infrastructure (grant~ek9) and the Pawsey Supercomputing Centre (project~pawsey0810) in the framework of the National Computational Merit Allocation Scheme and the ANU Merit Allocation Scheme. The simulation software, \texttt{FLASH}, was in part developed by the Flash Centre for Computational Science at the University of Chicago and the Department of Physics and Astronomy at the University of Rochester.
\\

\noindent
The authors report no conflict of interest.

\appendix
\section{Simulation parameters}
\label{appen:numerical}
For completeness, we list details of all simulation runs that were used in the analysis. More details can be found in \citet{AFS2026}. This includes various tests and validations of the statistics in the subsonic and supersonic regimes. Tables~\ref{tab:tab1} and~\ref{tab:tab1a} list the simulation parameters for all DNS considered in this study, together with the resolution measure $k_{\rm max}\eta$ and the number $N_s$ of statistically independent snapshots used for averaging. Successive snapshots are sampled at intervals of $T_e/4$, where $T_e=L_f/u_{\rm rms}$ is the eddy-turnover time.

\begin{table*}
\setlength{\tabcolsep}{10pt}
\centering
\renewcommand{\arraystretch}{0.9}
\begin{tabular}{ccccccc}
\hline \hline\\[-7pt]
${\rm Run}$ & $N^3$   & $M_{\rm rms}$  & $Re$ & $Re_\lambda$ & $N_s$ &  $k_{\rm max} \eta$ \\[3pt]
\hline \\[-6pt]
\quad $1$  & $512^3$  & $0.10$  & $100$ & $17$ & $100$  & $29.17$ \\
\quad $2$  & $512^3$  & $0.10$  &  $219$& $29$  & $100$  & $17.45$\\
\quad $3$  & $512^3$  & $0.10$ &  $399$& $43$ & $100$  & $11.58$\\
\quad $4$  & $1024^3$  & $0.10$ & $520$ & $50$ & $100$  & $19.24$ \\
\quad $5$  & $1024^3$  & $0.10$ & $709$ & $58$ & $100$  & $15.26$ \\
\quad $6$  & $1024^3$  & $0.10$ & $932$ & $67$ & $100$ & $12.40$ \\	
\quad $7$  & $2048^3$  & $0.10$ & $2469$& $117$ & $100$ & $12.40$ \\[5pt]
\quad $8$  & $512^3$  & $0.34$  & $102$ & $17$  & $100$ & $29.06$  \\
\quad $9$  & $512^3$ & $0.33$ & $222$  & $29$ & $100$   & $17.29$ \\
\quad $10$  & $512^3$ & $0.33$  & $391$ & $42$  & $100$ & $11.74$ \\
\quad $11$  & $1024^3$ & $0.33$ & $511$  & $49$ & $100$   & $19.46$ \\
\quad $12$  & $1024^3$ & $0.33$ & $715$  & $59$ & $100$   & $15.26$ \\
\quad $13$  & $1024^3$  & $0.33$ & $956$  & $70$ & $100$  & $12.39$ \\
\quad $14$  & $2048^3$ & $0.33$ & $2411$ & $116$ & $100$  & $12.64$  \\[5pt]
\quad $15$  & $512^3$  & $0.55$ & $96$ & $16$ & $200$  & $29.88$ \\
\quad $16$  & $512^3$  & $0.54$ & $213$ & $30$ & $200$  & $18.33$ \\
\quad $17$  & $512^3$  & $0.54$ & $378$ & $41$ & $200$  & $12.05$ \\
\quad $18$  & $1024^3$  & $0.55$ & $497$ & $48$  & $200$  & $19.75$ \\
\quad $19$  & $1024^3$  & $0.56$ & $712$ & $61$ & $100$  & $15.51$ \\
\quad $20$  & $1024^3$  & $0.56$  & $955$ & $71$ & $100$  & $12.46$ \\
\quad $21$  & $2048^3$  & $0.56$ & $2405$ & $115$ & $100$  & $12.60$  \\[5pt]
\quad $22$  & $512^3$  & $0.78$  & $97$ & $16$ & $200$  & $29.55$ \\
\quad $23$ & $512^3$  & $0.77$  & $212$ & $29$ & $200$  & $17.91$  \\
\quad $24$  & $512^3$  & $0.77$  & $385$ & $42$ & $200$  & $11.91$ \\
\quad $25$ & $1024^3$  & $0.78$  & $505$ & $49$ & $200$  & $19.68$ \\
\quad $26$ & $1024^3$  & $0.77$  & $695$ & $59$ & $100$  & $15.68$ \\
\quad $27$  & $1024^3$  & $0.78$ & $947$  & $70$ & $100$  & $12.51$ \\
\quad $28$  & $2048^3$  & $0.78$ & $2414$ & $116$ & $100$  & $12.62$  \\[5pt]
\quad $29$  & $512^3$  & $1.0$  & $95$ & $16$ & $200$  & $29.60$ \\
\quad $30$  & $512^3$  & $1.0$  & $218$ & $29$ & $200$  & $17.40$  \\
\quad $31$  & $512^3$  & $1.0$  & $386$ & $42$ & $200$  & $11.88$ \\
\quad $32$  & $1024^3$  & $1.0$  & $499$ & $49$ & $200$  & $19.78$ \\
\quad $33$  & $1024^3$  & $1.0$  & $686$ & $58$ & $100$  & $15.73$ \\
\quad $34$  & $1024^3$  & $1.0$ & $927$  & $69$ & $100$  & $12.64$ \\
\quad $35$  & $2048^3$  & $1.0$ & $2389$ & $115$ & $100$  & $12.69$  \\
\hline\hline
\end{tabular}
\vspace{0.5cm}
\caption{DNS parameters  of solenoidally forced compressible isotropic turbulence. The Run number, grid points $N^3$, rms Mach number $M_{\rm rms}$, large-scale Reynolds number $Re$, Taylor micro-scale Reynolds number $Re_{\lambda}$, statistically independent snapshots $N_s$, and $k_{\rm max}\eta$.}
\label{tab:tab1}
\end{table*}

\begin{table*}
\setlength{\tabcolsep}{10pt}
\centering
\renewcommand{\arraystretch}{0.9}
\begin{tabular}{cccccccccc}
\hline \hline\\[-7pt]
${\rm Run}$ & $N^3$   & $M_{\rm rms}$  & $Re$ & $Re_\lambda$ & $N_s$  & $k_{\rm max} \eta$ \\[3pt]
\hline \\[-6pt]
\quad $36$  & $512^3$  & $3.0$  & $95$ & $15$ & $200$  & $28.32$ \\
\quad $37$  & $512^3$  & $3.0$  & $217$ & $25$ & $200$ & $16.23$  \\
\quad $38$  & $512^3$  & $3.1$  & $390$ & $35$ & $200$  & $10.76$ \\
\quad $39$  & $1024^3$  & $3.0$  & $507$ & $40$ & $200$  & $17.72$ \\
\quad $40$  & $1024^3$  & $3.0$  & $721$ & $49$  & $200$ & $13.92$ \\
\quad $41$  & $1024^3$  & $3.1$ & $944$  & $59$ & $200$  & $11.31$ \\
\quad $42$  & $2048^3$  & $3.0$ & $2360$ & $96$ & $132$  & $11.74$  \\[5pt]
\quad $43$  & $512^3$  & $6.2$  & $104$ & $16$ & $200$ & $26.94$ \\
\quad $44$  & $512^3$  & $6.2$  & $207$ & $24$ & $200$  & $16.77$  \\
\quad $45$  & $512^3$  & $6.2$  & $416$ & $36$ & $200$  & $10.22$ \\
\quad $46$  & $1024^3$  & $6.2$  & $519$ & $41$ & $200$  & $17.35$ \\
\quad $47$  & $1024^3$  & $6.2$  & $724$ & $49$  & $200$  & $13.64$ \\
\quad $48$  & $1024^3$  & $6.2$ & $1034$  & $59$ & $200$ & $10.49$ \\
\quad $49$  & $2048^3$  & $6.3$ & $2631$ & $97$ & $150$ & $10.49$  \\[5pt]
\quad $50$  & $512^3$  & $10.2$  & $98$ & $15$ & $200$  & $28.42$ \\
\quad $51$  & $512^3$ & $10.2$  & $222$ & $25$ & $200$  & $16.01$  \\
\quad $52$  & $512^3$  & $10.3$  & $396$ & $36$ & $200$  & $10.64$ \\
\quad $53$  & $1024^3$  & $10.2$  & $510$ & $40$ & $200$  & $17.56$ \\
\quad $54$  & $1024^3$  & $10.2$  & $714$ & $48$ & $200$  & $13.69$ \\
\quad $55$  & $1024^3$  & $10.3$ & $958$  & $56$ & $200$  & $11.08$ \\
\quad $56$  & $2048^3$  & $10.4$ & $2445$ & $92$ & $146$  & $11.12$  \\
\hline\hline
\end{tabular}
\vspace{0.5cm}
\caption{Continued from Table I.}
\label{tab:tab1a}
\end{table*}

\def\apj{{\sl Astrophys.~J.}}
\def\apjl{{\sl Astrophys.~J.}}
\def\apjs{{\sl Astrophys.~J.~Suppl.~Ser.}}
\def\aap{{\sl Astron.~Astrophys.}}
\def\aaps{{\sl Astron.~Astrophys.~Suppl.~Ser.}}
\def\aj{{\sl Astron.~J.}}
\def\araa{{\sl Annu.~Rev.~Astron.~Astrophys.}}
\def\mnras{{\sl Mon.~Not.~R.~Astron.~Soc.}}
\def\physrep{{\sl Phys.~Rep.}}
\def\prl{{\sl Phys.~Rev.~Lett.}}
\def\jfm{{\sl J.~Fluid Mech.}}
\def\rmp{{\sl Rev.~Mod.~Phys.}}
\def\jcp{{\sl J.~Comput.~Phys.}}

\bibliographystyle{jfm}
\bibliography{all,federrath}

\end{document}